\documentclass[aps,prd,preprintnumbers,nofootinbib,onecolumn,superscriptaddress]{revtex4-2}
\usepackage{amsmath,amssymb}
\usepackage[dvipdfmx]{graphicx}
\usepackage{float}
\usepackage{comment}
\usepackage{xcolor}

\usepackage{bm}

\begin{document}


\newcommand{\simgt}{\lower.5ex\hbox{$\; \buildrel > \over \sim \;$}}
\newcommand{\simlt}{\lower.5ex\hbox{$\; \buildrel < \over \sim \;$}}

\title{Large-scale structure with superhorizon isocurvature dark energy}
\author{Koki Yamashita}
\affiliation{
Department of Physics, Kyushu University, Motooka 744, Fukuoka, 819-0395 Japan}
\author{Yue Nan}
\email{yue.nan@ipmu.jp}
\affiliation{
Physics Program, Graduate School of Advanced Science and Engineering, Hiroshima University, 1-3-1 Kagamiyama, Higashi-hiroshima, 739-8526, Japan}
\affiliation{Kavli Institute for the Physics and Mathematics of the Universe (WPI),
The University of Tokyo Institutes for Advanced Study,
The University of Tokyo, Kashiwa, Chiba 277-8583, Japan}
\author{Yuuki Sugiyama}
\email{sugiyama.yuki@phys.kyushu-u.ac.jp}
\affiliation{
Department of Physics, Kyushu University, Motooka 744, Fukuoka, 819-0395 Japan}
\author{Kazuhiro Yamamoto}
\email{yamamoto@phys.kyushu-u.ac.jp}
\affiliation{
Department of Physics, Kyushu University, Motooka 744, Fukuoka, 819-0395 Japan}
\affiliation{Research Center for Advanced Particle Physics, Kyushu University, 744 Motooka, Nishi-ku,  Fukuoka 819-0395, Japan}
\begin{abstract}
The standard cosmological model assumes a homogeneous and isotropic universe as the background spacetime on large scales called the cosmological principle. However, some observations suggest the possibility of an inhomogeneous and anisotropic universe at large scales. 
In this paper, we investigate a model of the universe with random inhomogeneities and anisotropies on very large scales, motivated by the supercurvature dark energy model in Nan \textit{et al.} [Phys. Rev. D 99, 103512 (2019)]. In this model, the authors introduced a scalar field with $\mathcal{O}(1)$ inhomogeneities on a scale sufficiently larger than the current horizon scale (superhorizon scale), and the potential energy of the scalar field explains the accelerating expansion, with slight deviations from the cosmological principle. We aim at clarifying the theoretical prediction on the large-scale structure (LSS) of the matter component in this model.
Based on the work on the superhorizon scale fluctuations (superhorizon mode) presented in Nan and Yamamoto [Phys. Rev. D 105, 063518 (2022)], we derive the equations that the perturbative components to the LSS obey as a generalization of the cosmological perturbations theory, which is
solved to find the influence of the dark energy inhomogeneities on the formation of the LSS. Finally, we show that the model can be 
consistent with observations by comparing the $\sigma_8$ predicted by the numerical solution of the model with the $\sigma_8$ indicated by observations such as Planck and the Sloan Digital Sky Survey.
\end{abstract}
\maketitle

\section{\label{sec:level1}Introduction}
Currently, observations of Type Ia supernovae and the cosmic microwave background (CMB) have confirmed that we live in an acceleratively expanding universe. It is known that the expansion rate of the Universe changes from decelerating to accelerating at a particular redshift called the transition redshift, but the actual cause of this phase transition remains unknown. To explain the present universe, models assuming the existence of an energy source that lives with general relativity (GR) or that modifies GR are being considered. With GR, the current accelerating expansion of the Universe is generally driven by some exotic energy called dark energy. Since the late 1990s, this dark energy has become an essential topic in cosmology, in addition to cold dark matter (CDM)~\cite{ade2014planck,aghanim2020planck}, and is known to account for about 70$\%$ of the total energy density of the Universe today. The simplest and most well-accepted model to explain the accelerated expansion, the cosmological constant $\Lambda$ model as dark energy, is consistent with extensive observations and is still considered the leading one, which adds to CDM to build up the $\Lambda$CDM standard cosmological model.

However, while the standard cosmological model assumes isotropy and homogeneity on large scales, several studies suggest that there may be deviations from the cosmological principle generally assumed in the standard cosmological model. Some examples are as follows. (i)  {\it Hemispherical asymmetry of CMB power spectrum.}~The observed CMB power spectrum suggests the existence of a power asymmetry modulated as a dipole, analyzed from data of the Wilkinson Microwave Anisotropy Probe (WMAP) of different hemispheres over the sky in Ref.~\cite{eriksen2007hemispherical}, and the Planck results also confirm this dipole modulation ~\cite{aghanim2014planck,Ghosh_2016}. Recently, the authors of Ref.~\cite{Secrest_2021} reported that the observed dipole of the sky of Quasi-stellar objects (QSO) rejects simply following the canonical, kinematic Doppler interpretation of the CMB dipole(cf. Ref.~\cite{Yamamoto:2003vh}).
(ii) {\it Directional dependence of Hubble parameter $H_0$.}~The investigation on the Hubble constant over the entire sky using the x-ray luminosity-temperature relation of galaxy clusters suggests a directional dependence of the expansion rate, which indicates that the expansion rate of the local Universe may be anisotropic~\cite{migkas}; this possibility of anisotropic $H_0$ is reinforced by analyses from 10 different scaling relations of galaxy clusters additional to the luminosity-temperature relation~\cite{Migkas:2021zdo}. Moreover, similar directional dependence of $H_0$ is suggested by observations on QSO and gamma-ray burst sources~\cite{luongo2021larger}.
Additionally, other possible deviations have been observed (see, e.g., Ref.~\cite{perivolaropoulos2014large} for an extensive review on large-scale anomalies, Ref.~\cite{rath2013testing} for possible scale dependence of the dipole modulation, Ref.~\cite{muir2018covariance} for low-multipole alignments, and Ref.~\cite{schwarz2016cmb} for a comprehensive review on CMB anomalies). Out of these anomalies in observations, the possibility of probing the breakdown of cosmological principle is being examined by some authors~\cite{PhysRevD.105.063514}. 
In addition, there is a recent work testing the cosmological principle based on the $\Lambda$CDM model with the CMASS galaxy sample of BOSS DR12~\cite{Kim:2021osl}.

Since several previous observations and analyses suggest possible deviations from the cosmological principle, it is worthwhile enough to consider a cosmological model with large-scale inhomogeneity that violates the cosmological principle. Our motivation is to reveal whether a model with large-scale inhomogeneity is consistent with observations and to figure out the theoretical predictions of the model. In particular, this paper investigates the effect of inhomogeneity on the large-scale structure (LSS) of the Universe.

A commonly used statistical measurement for the density perturbations in the Universe is the power spectrum $P(k)$. 
Recent observations on the power spectrum of the LSS 
have been accurately performed by galaxy surveys such as the Sloan Digital 
Sky Survey (SDSS) and the Dark Energy Survey (DES). 
Therefore, the purposes of this research are to determine the power spectrum $P(k)$, to find out how the large-scale nonuniformity affects it, to determine the order of magnitude of the effect, and to check the consistency with observations.

In recent years, cosmological models breaking the cosmological principle have been constructed to study various aspects of the accelerating Universe \cite{yadav2012bianchi,amirhashchi2017viscous,amirhashchi2019current,amirhashchi2018probing,mishra2017dark,mishra2019bulk}. In this paper, we consider a simplified version of the supercurvature mode dark energy model~\cite{nan2019large}, which is a stochastic model of dark energy with large-scale inhomogeneity assuming an open universe associated with a specific inflationary scenario.
The model introduces the potential energy of a scalar field with fluctuations of $\mathcal{O}(1)$ on a supercurvature scale sufficiently larger than the current horizon scale, which is responsible for the accelerated expansion. 
Motivated by the model, in the present paper, we consider the dark energy model of a scalar field with inhomogeneities on scales larger than the current horizon scale assuming the spatial curvature $K$ is set to $K=0$ . In the following, we first review the previous study \cite{nan2021dark} to introduce this model, and then find the equations governing the terms of perturbations relevant to the LSS, the solutions for the evolution of the perturbations, and the formulation for the power spectrum based on these solutions.

The remaining parts of the paper are organized as follows. 
In Sec.~\ref{sec:level2}, we explain the definitions for the perturbations introduced by the large-scale inhomogeneity of the dark energy and introduce the basic setups for the formulation.
In Sec.~\ref{sec:level3}, specific derivations leading to the power spectrum of the LSS of matter following the formulation are performed.
Section~\ref{sec:level4} is devoted to the details of numerical calculation aimed at quantitative evaluation of the modifications in matter distribution introduced by the inhomogeneous model.
In Sec.~\ref{sec:level5}, we conclude the results and discuss their implications on the cosmological model.
The Appendix is attached as additional explanations for parts of the formulations and is organized as follows.
In Appendix~\ref{AppendixA}, we review the derivation of the Einstein equations following the metric perturbations from the superhorizon inhomogeneities of dark energy introduced by this model.
Appendix~\ref{AppendixC} presents a generalized expression for higher order of a source term concerning the solution of the equation that governs the isocurvature mode,  which is introduced in Sec.~\ref{subsec:formulation}.
Appendix~\ref{AppendixB} follows as additional details of the analytic approximations for the modification to the LSS power spectrum, which arises from different modes of perturbations induced by large-scale inhomogeneity of dark energy introduced in this paper; the approximations are useful in the numerical evaluation and helpful for comprehending the behaviors of the modifications induced by our model.
Throughout the present paper, we use the unit that the velocity of light equals unity $c=1$, and $M_{pl}$ denotes the Planck mass defined by $M_{pl}^2=(8\pi G)^{-1}$, where $G$
is the gravitational constant.

\section{\label{sec:level2}Basis of the formulation}

We formulate for the evolution of matter density fluctuation with  
dark energy inhomogeneities on superhorizon scales based on the standard perturbation theory. Subsequently, we evaluate the model predictions on the matter power spectrum with the solution of the system. We are mainly interested in the LSS formation in the late time, where the modification on observables such as the matter power spectrum are expected to occur. For the late-time evolution of cosmological perturbations, in addition to the cold dark matter, we consider that the inhomogeneous dark energy on superhorizon scales is sourced by some light scalar field $\phi$, which is also treated as perturbations to a homogeneous and isotropic universe. We start with the basic equations governing the evolution of the dark matter, the scalar field $\phi$,  and the metric perturbations.

This section shows how to describe different modes of perturbations and how to derive the equations that the components of perturbations follow. The formation of LSS of galaxies in a uniformly isotropic universe has already been well known within the cosmological perturbation theory. We start with generalizing the framework of cosmological perturbation theory in the homogeneous universe to that in an inhomogeneous universe.
\subsection{Basic equations}
In the following, the basic equations in the framework of cosmological perturbation theory are introduced. Following linear perturbation theory, we adopt the conformal Newtonian gauge,
\begin{align}
ds^2=-(1+2\Psi(t,\bm x))dt^2+a^2(t)(1 + 2 \Phi(t,\bm{x}))dx^idx^j\delta_{ij}.
\label{ds}
\end{align}
As for the metric perturbations, we consider the equations up to the first order of  $\Psi$ and $\Phi$.
The Einstein equation associated with the energy-momentum tensor is
\begin{align}
\label{einstein}
G^{\mu}_{\  \nu} = 8\pi G(T^{\mu\  (m)}_{\  \nu} + T^{\mu\  (\phi)}_{\  \nu} ).
\end{align}
Here $T^{\mu \  (m)}_{\  \nu}$ and $T^{\mu \  (\phi)}_{\  \nu}$ are the energy-momentum tensor of the matter and the scalar field, respectively, 
which follow the conservation equation $\nabla_\mu T^\mu{}_\nu^{(c)}=0$ with 
$c=m ~({\rm matter~component}),~\phi~({\rm scalar~field~component})$. 
The equation of continuity for the matter component in an expanding universe reads
\begin{align}
\label{con}
\frac{\partial \rho}{\partial t} + 3H\rho + \frac{1}{a}\partial_i(\rho v^i) + 3\dot{\Phi}\rho = 0,
\end{align}
and the Euler equation reads,
\begin{align}
\label{eul}
\frac{\partial}{\partial t}(\rho v^i) + 4H\rho v^i + \frac{1}{a}\partial_j(\rho v^i v^j) +\frac{\rho}{a} \frac{\partial \Psi}{\partial x^i} + 4\dot{\Phi}\rho v^i=0.
\end{align}

The Klein-Gordon equation for the scalar field denoted by $\phi$ is
\begin{align}
\label{klein-gordon}
\frac{1}{\sqrt{-g}}\partial_{\mu} (\sqrt{-g}  g^{\mu \nu}\partial_{\nu} \phi) -m^2\phi = 0.
\end{align}
As assumed in the previous works \cite{nan2019large,nan2021dark}, we consider the model that the ultralight scalar field is varying at a very large spatial scale, much longer than the Hubble length in the present Universe. Thus, they create large-scale inhomogeneities of the dark energy on superhorizon scales and break the cosmological principle, but the violation is small within the horizon of our Universe at present. To describe the LSS in our model, we introduce two parameters. 
One is the parameter  $\epsilon$ to describe the inhomogeneities on the superhorizon scales coming from the ultralight scalar field $\phi$. The other is the parameter $\kappa$ for the standard cosmological perturbations. 
Our approach is based on the scheme with two-parameter expansion.

\subsection{\label{sec:citeref}Perturbations in an inhomogeneous universe}
This subsection describes the notations for the perturbations to clarify the LSS with the inhomogeneities on the superhorizon scales. First, as for the metric perturbations, we consider the first-order perturbation of $\Psi$ and $\Phi$ in Eq.~(\ref{ds}). Furthermore, we introduce the two expansion parameters $\epsilon$ and $\kappa$. Within them, $\epsilon$ describes the perturbations of superhorizon scales, whereas $\kappa$ describes the conventional cosmological perturbations of the LSS of galaxies on scales well inside the horizon scale. 
We denote the superhorizon perturbations (superhorizon modes, for short, sh-mode) by quantities labeled with ``sh" and the perturbations to the LSS by quantities labeled with ``lss". 
Then the metric perturbations can be written as 
\begin{align}
\Phi &= \kappa\ ^{\mathrm{lss}}\Phi+ \epsilon \ ^{\mathrm{sh}}\Phi,\\
\Psi  &= \kappa\ ^{\mathrm{lss}}\Psi+ \epsilon \ ^{\mathrm{sh}}\Psi.
\end{align}
Similarly, as for the density perturbations and the velocity field, we write  
\begin{align}
    &\rho=\rho_0(t)(1+\kappa\ ^{\mathrm{lss}}\delta + \epsilon \ ^{\mathrm{sh}}\delta),
    \\
    &v^i = \kappa\ ^{\mathrm{lss}}v^i + \epsilon \ ^{\mathrm{sh}}v^i.
\end{align}
The scalar field has only the superhorizon scale perturbations; then, we write
\begin{align}
\label{phi-exp}
\phi &= \phi_0 + \epsilon\ ^{\mathrm{sh}}\phi, 
\end{align}
where $\phi_0$ is the dark energy background. 
The evolution of the superhorizon perturbations of $\phi$ was investigated in Ref.~\cite{nan2021dark}, which is briefly reviewed in Sec. \ref{subsec:sh-mode}.
In the present paper, we focus on investigating the effect of the superhorizon inhomogeneities on the LSS perturbations, which is described by the perturbations of the order of ${\cal O}(\kappa\epsilon)$. 

The superhorizon perturbations of the scalar field behave as an isocurvature inhomogeneous dark energy model, 
which give rise to the perturbations of the order of  ${\cal O}(\kappa\epsilon)$ through the coupling with the standard perturbations on the LSS.
Consequently, to characterize the effect of the order of ${\cal O}(\kappa\epsilon)$, we may write
\begin{align}
\kappa \ ^{\mathrm{lss}}\delta \equiv \kappa (\ ^{\mathrm{ad}}\delta +\epsilon\ ^{\mathrm{iso}}\delta).
\end{align}
In the following, we ignore the terms higher order than $\mathcal{O}(\kappa^a\epsilon^b)$ with $a>1$ and $b>1$ and keep the coupling terms of $\mathcal{O}(\kappa^1 \epsilon^1)$ which describes the imprints that the sh-mode induces on the LSS matter distribution.
Correspondingly, we assume the following form for the perturbation expansion in addition to Eq.~(\ref{phi-exp}), 
\begin{align}
\label{notationa}
&\rho=\rho_0(1 + \kappa\ ^{\mathrm{lss}}\delta+\epsilon\ ^{\mathrm{sh}}\delta)
=\rho_0(1+\kappa\ ^{\mathrm{ad}}\delta) + \epsilon(\ ^{\mathrm{sh}}\rho + \kappa\rho_0 \ ^{\mathrm{iso}}\delta), \\
\label{notationb}
&v^i = \kappa\ ^{\mathrm{ad}}v^{i} + \epsilon(\ ^{\mathrm{sh}}v^i + \kappa\ ^{\mathrm{iso}}v^{i}), \\
\label{notationc}
&\Phi=\kappa \ ^{\mathrm{ad}}\Phi + \epsilon (\ ^{\mathrm{sh}}\Phi+ \kappa\ ^{\mathrm{iso}}\Phi), \\
\label{notationd}
&\Psi = \kappa\ ^{\mathrm{ad}}\Psi + \epsilon (\ ^{\mathrm{sh}}\Psi + \kappa\ ^{\mathrm{iso}}\Psi), 
\end{align}
where $^{\mathrm{sh}}\rho \equiv \rho_0 \ ^{\mathrm{sh}}\delta$. 
We note that the quantities labeled with ``ad'' describe the conventional adiabatic perturbations 
of the LSS, while the quantities labeled with ``iso'' describe the perturbations
on the LSS generated by the coupling with the superhorizon inhomogeneities of dark energy, whose effect is negligible in the early time hence obeying the isocurvature initial condition.

To summarize this subsection, the evolution of the superhorizon mode was expanded and represented by $\epsilon$, 
which was investigated in Ref.~\cite{nan2019large}. 
In the following, after briefly reviewing the background equations and the superhorizon perturbations of the order  $\kappa$, 
we find the expression for the perturbation to the LSS of the order $\kappa\epsilon$  as an original work of the present paper.

\subsection{The homogeneous background [$\mathcal{O}(\kappa^0\epsilon^0)$]}
We first consider the homogeneous background evolution of the model of the order $\mathcal{O}(\kappa^0\epsilon^0)$, which is well known, without the inhomogeneities induced by the lss-mode and the sh-mode perturbations introduced in Sec.~\ref{sec:citeref}. From the fluid equation in an expanding universe, the equation of motion of the scalar field (Klein-Gordon equation), and the Einstein equation,  the background equations follow:
\begin{gather}
\label{background1}
\dot{\rho_0} + 3H\rho_0 = 0,\\
\label{background2}
\ddot{\phi}_0(t) +3H(t)\dot{\phi}_0(t)  +m^2\phi_0(t) = 0, \\
\label{background3}
3H^2 = 8\pi G(\frac{1}{2}\dot{\phi}_0^2 + \frac{1}{2}m^2\phi_0^2  +\rho_0),\\
\label{background4}
(2\frac{\ddot{a}}{a}+H^2) = 8\pi G ( \frac{1}{2}m^2\phi_0^2-\frac{1}{2}\dot{\phi}_0^2 ).
\end{gather}
Here, $\rho_0$ is the background density, and $m$ is the scalar field mass.

\subsection{\label{subsec:sh-mode}Superhorizon modes [$\mathcal{O}(\kappa^0\epsilon^1)$]}
We here review the evolutions of the superhorizon mode (sh-mode) perturbations of the order $\mathcal{O}(\kappa^0\epsilon^1)$. 
First, the sh-mode can be spatially expanded and rewritten as the following multipolar form; for instance, for $\phi$, we have
\begin{align}
\epsilon ^{\mathrm{sh}}\phi(t,\bm{x}) &= \epsilon\sum_{n=1}^3\phi^{(n)}_1(t)T_i^{(n)}x^i 
+\mathcal{O}(\epsilon^2 x^2),
 \label{multipole}
\end{align}
where $T_i^{(n)}$ is defined by
\begin{align}
  T^{(n=1)}_{i} = \sqrt{\frac{3}{4\pi}}
    \left( 
      \begin{array}{ccc}
       1 \\
       0 \\
       0
      \end{array} 
      \right)
      \qquad \qquad
  T^{(n=2)}_{i} = \sqrt{\frac{3}{4\pi}}
    \left( 
      \begin{array}{ccc}
        0 \\
        1 \\
        0
       \end{array} 
      \right)
      \qquad \qquad
  T^{(n=3)}_{i} = \sqrt{\frac{3}{4\pi}}
    \left( 
     \begin{array}{ccc}
      0 \\
      0 \\
      1
     \end{array} 
    \right),
          \label{defT}
\end{align} 
and the subscript ``$1$'' in $\phi_1^{(n)}(t)$ denotes
the dipole component. 
Throughout the present paper, we implicitly assume the Einstein summation convention with respect to the index of $x^i$. 
Similarly, for the sh-mode up to the order of $\mathcal{O}(\kappa^0\epsilon^1)$, 
we can expand 
\begin{align}
&\ ^{\mathrm{sh}}v^i\equiv \partial_i\ ^{\mathrm{sh}}V
\simeq \sum_{n=1}^3V^{(n)}_1(t)T_i^{(n)},
\label{sh-v-dipole}\\ 
&\ ^{\mathrm{sh}}\delta \simeq\sum_{n=1}^3\delta^{(n)}_1(t)T_i^{(n)}x^i,
\label{sh-delta-dipole}\\
&\ ^{\mathrm{sh}}\Phi\simeq \sum_{n=1}^3\Phi_1^{(n)}(t)T_i^{(n)} x^i,
\label{sh-Phi-dipole}
\end{align}
where we introduced the velocity potential ${}^{\rm sh}V= \sum_{n=1}^3V^{(n)}_1(t)T_j^{(n)}x^j$, which is associated with the sh-mode velocity. 
From Eq.~(\ref{sh-v-dipole}), it is worth mentioning that $\ ^{\mathrm{sh}}v^i$ is spatially uniform and plays a role similar to the dipole modulation of the background density.

Next, we introduce the equations of the sh-mode, which were given in Ref.~\cite{nan2021dark}, and a review with more details of this part given in Appendix~\ref{AppendixA}. 
The sh-mode of $\mathcal{O}(\kappa^0\epsilon^1)$  follows the equations
\begin{gather}
\label{sh1}
\delta^{(n)}_{1} +3\Phi^{(n)}_{1}=0,\\
\label{sh3}
\dot{V}'^{(n)}_{1} - \Psi^{(n)}_{1} =0,\\
\label{sh4}
\ddot{\phi}^{(n)}_{1}+ 3H\dot{\phi}^{(n)}_{1} + m^2 \phi^{(n)}_{1} + \left(3\dot{\Phi}^{(n)}_{1} - \dot{\Psi}^{(n)}_{1}- 6H\Psi^{(n)}_{1}\right)\dot{\phi}_0 - 2\Psi^{(n)}_{1}\ddot{\phi}_0 =0,\\
\label{sh5}
3H\left(\dot{\Phi}^{(n)}_{1} - H\Psi^{(n)}_{1}\right) = 4\pi G( \rho_0\delta^{(n)}_{1} + m^2\phi_0\phi_{1}-\dot{\phi}^2_0 \Psi^{(n)}_{1} +\dot{\phi}_0\dot{\phi}^{(n)}_{1}  ),\\
\label{sh6}
\dot{\Phi}^{(n)}_{1} -H\Psi^{(n)}_{1}  = -4\pi G(\rho_0V'^{(n)}_{1} +\dot{\phi}_0\phi^{(n)}_{1}),\\
\label{sh2}
\Psi^{(n)}_{1} + \Phi^{(n)}_{1} = 0,
\end{gather}
where all subscripts  ``$1$'' stand for the dipole 
of the perturbations, 
and we introduced $V'^{(n)}_1\equiv -aV^{(n)}_1$.\footnote{
To clarify the difference of the notations between this paper and the previous work Ref.~\cite{nan2021dark}, 
we note that in this paper, $V'^{(n)}_1$ is equivalent with the $V_{1(m)}$ defined in Sec. II A of Ref.~\cite{nan2021dark} instead of $V_1$.
The behaviors of the background $\phi_0$ and dipole perturbations $\phi_1$, $\Phi_1$ and $V_1$ predicted by the model under different parameters can be found in Ref.~\cite{nan2021dark}.}
We also note that the initial condition for the sh-mode perturbation is the isocurvature type; hence, excepting $\phi_1^{(n)}$, we assume that $\delta_1^{(n)}=\Psi_1^{(n)}=\Phi_1^{(n)}=V_1'^{(n)}=0$ at the initial time. The initial value of $\phi_1^{(n)}$ determines the amplitude of the superhorizon mode perturbations
(see Appendix \ref{AppendixB}).
The evolution of the superhorizon modes was investigated in Ref.~\cite{nan2021dark} using both an analytic method and a numerical method based on the above equations. 
We remark on some of the results. The superhorizon modes come from the scalar field perturbations on 
the superhorizon scales, which is understood as an inhomogeneous dark energy model effectively. 
The matter density perturbations in the region of the higher amplitude of the scalar field decrease in comparison with other regions because the larger amplitude of the dark energy makes the Universe expand faster and suppresses 
the growth of the density perturbations. Then, the sign of the density perturbation of the sh-modes 
becomes opposite to that of the scalar field perturbation. 

We can put a constraint on the sh-mode using the dipole observation in the CMB, as is discussed in Sec. \ref{observationalconstraint} and in Ref. \cite{nan2021dark}. 
Our model predicts dipole anisotropies in the CMB due to the sh-modes. 
The dipole observations in the CMB put a constraint on the predictions of our model.
We roughly find a constraint on the order of magnitude of the sh-mode as 
$\epsilon \lesssim \mathcal{O}(10^{-2})$.
Usually, the CMB dipole is explained by the Doppler effect due to the Galaxy motion, 
but it can also be explained by the inhomogeneity of the sh-modes in our model. 
A combination of these two effects may explain the CMB dipole. 
The constraint on the sh-mode $\epsilon \lesssim \mathcal{O}(10^{-2})$
comes from the assumption that the CMB dipole 
cannot exceed the prediction caused by the sh-modes.
A more accurate value of the constraint is given later using the Planck data.

\section{\label{sec:level3}Specific formulation of the LSS [$\mathcal{O}(\kappa^1\epsilon^1)$]}
In our following formulation, for clarity, we temporarily write the terms related to $\kappa$ denoting the ad-mode/lss-mode and $\epsilon$ denoting the iso-mode/sh-mode explained earlier in Sec.~\ref{sec:citeref} and keep them to the order of $\mathcal{O}(\kappa^1\epsilon^1)$ that we are interested in.
Thereafter, we concretely seek the evolution of the density fluctuations in a nonuniform universe, described by Eqs. (\ref{con}) and (\ref{eul}). We will perform the calculation based on the argument of Sec.~\ref{sec:level2} in the following steps: (i) Step 1, we substitute  Eqs.~(\ref{notationa})-(\ref{notationd}) into Eqs. (\ref{con}) and (\ref{eul}). (ii) Step 2, we introduce Fourier transformation with the first-order perturbation expansion and obtain the two equations. (iii) Step 3, we obtain formulations to describe the matter density perturbations 
and the matter power spectrum under the influence of 
inhomogeneous dark energy.

\subsection{Substitution of different modes into the fluid equations}
First, we substitute Eqs.~(\ref{notationa})-- (\ref{notationd}) into the continuity Eq.~(\ref{con}); then, we have
\begin{align}
\label{cont-shlss}
& \dot{\rho_0}(1+\kappa\ ^{\mathrm{ad}}\delta) +\kappa \rho_0 \ ^{\mathrm{ad}}\dot{\delta} + \epsilon(\ ^{\mathrm{sh}}\dot{\rho} + \kappa\dot{\rho_0} \ ^{\mathrm{iso}}\delta + \kappa\rho_0 \ ^{\mathrm{iso}}\dot{\delta}) 
\nonumber\\
&\quad +3H(\rho_0(1+\kappa\ ^{\mathrm{ad}}\delta) + \epsilon(\ ^{\mathrm{sh}}\rho + \kappa\rho_0 \ ^{\mathrm{iso}}\delta)) 
\nonumber\\
&\quad + \frac{1}{a}\partial_i\Big\{\rho_0(1+\kappa\ ^{\mathrm{ad}}\delta)\kappa\ ^{\mathrm{ad}}v^{i} + \epsilon\left[\rho_0(1+\kappa\ ^{\mathrm{ad}}\delta)(\ ^{\mathrm{sh}}v^i + \kappa\ ^{\mathrm{iso}}v^{i})+(\ ^{\mathrm{sh}}\rho + \kappa\rho_0 \ ^{\mathrm{iso}}\delta)\kappa \ ^{\mathrm{ad}}v^{i}\right]\Big\} 
\nonumber\\
&\quad + 3\left\{ \kappa\ ^{\mathrm{ad}}\dot{\Phi}\rho_0(1+\kappa\ ^{\mathrm{ad}}\delta)  + \epsilon \left[\rho_0(1+\kappa\ ^{\mathrm{ad}}\delta) (\ ^{\mathrm{sh}}\dot{\Phi} + \kappa\ ^{\mathrm{iso}}\dot{\Phi})+ \kappa\ ^{\mathrm{ad}}\dot{\Phi}(\ ^{\mathrm{sh}}\rho + \kappa\rho_0 \ ^{\mathrm{iso}}\delta)\right]\right\}=0.
\end{align}
The continuity equation for the background density $\rho_0$ in a homogeneous expanding universe of the order $(\kappa^0\epsilon^0)$ reads
\begin{align}
\label{eq:conserve}
\dot{\rho_0} = -3H\rho_0,
\end{align}
which can be used to eliminate $\dot{\rho}_0$ in the following. 
Next, we consider the equations of perturbations. By inserting Eq.~(\ref{eq:conserve}) into Eq.~(\ref{cont-shlss}), the zeroth-order part of $\epsilon$, i.e., the $\mathcal{O}(\epsilon^0)$ piece, is
\begin{align}
 \label{eq:conti-0th}
\kappa\ ^{\mathrm{ad}}\dot{\delta} + \frac{1}{a}\partial_i\left\{(1+\kappa\ ^{\mathrm{ad}}\delta)\kappa\ ^{\mathrm{ad}}v^{i}\right\} + 3 \kappa\ ^{\mathrm{ad}}\dot{\Phi}(1+\kappa\ ^{\mathrm{ad}}\delta) =0,
\end{align}
while the first-order part of $\epsilon$, i.e., the $\mathcal{O}(\epsilon^1)$ piece, reads
\begin{align}
 \label{eq:conti-1st}
&\ ^{\mathrm{sh}}\dot{\rho}  + \kappa\rho_0 \ ^{\mathrm{iso}}\dot{\delta} + 3H\ ^{\mathrm{sh}}\rho  + \frac{1}{a}\partial_i\left[\rho_0(1+\kappa\ ^{\mathrm{ad}}\delta)(\ ^{\mathrm{sh}}v^i + \kappa\ ^{\mathrm{iso}}v^{i})+(\ ^{\mathrm{sh}}\rho + \kappa\rho_0 \ ^{\mathrm{iso}}\delta) \kappa\ ^{\mathrm{ad}}v^{i})\right] \nonumber \\
&\quad + 3\rho_0(1+\kappa\ ^{\mathrm{ad}}\delta) (\ ^{\mathrm{sh}}\dot{\Phi} + \kappa\ ^{\mathrm{iso}}\dot{\Phi})+ 3\kappa\ ^{\mathrm{ad}}\dot{\Phi}(\ ^{\mathrm{sh}}\rho + \kappa\rho_0 \ ^{\mathrm{iso}}\delta)=0.
\end{align}
Equation (\ref{eq:conti-0th}) is the familiar equation for the matter
adiabatic perturbations. 
The sh-mode follows the equation of the [$\mathcal{O}(\epsilon^1)$] in Eq.~(\ref{eq:conti-1st}),
\begin{align}
\label{sh_con}
\ ^{\mathrm{sh}}\dot{\rho} + 3H\ ^{\mathrm{sh}}\rho + 3\rho_0\ ^{\mathrm{sh}}\dot{\Phi}=0.
\end{align}
Here, we note that the $\partial_i\ ^{\mathrm{sh}}v^i$ term vanishes due to the spatial uniformity of ${}^{\rm sh}v^{i}$ in Eq.~(\ref{sh-v-dipole}). 
On the other hand, the remaining part of Eq.~(\ref{eq:conti-1st}) is
\begin{align}
\kappa \rho_0 \ ^{\mathrm{iso}}\dot{\delta}
+ 3\kappa\rho_0\ ^{\mathrm{ad}}\delta\ ^{\mathrm{sh}}\dot{\Phi} 
+ \frac{1}{a}\partial_i\left[\rho_0(\kappa+\kappa^2\ ^{\mathrm{ad}}\delta)\ ^{\mathrm{iso}}v^{i} + \kappa^2\rho_0\ ^{\mathrm{iso}}\delta\ ^{\mathrm{ad}}v^{i} + \rho_0(1+\kappa\ ^{\mathrm{ad}}\delta)\ ^{\mathrm{sh}}v^i  + \ ^{\mathrm{sh}}\rho \kappa\ ^{\mathrm{ad}}v^{i})\right] 
 &=0.
\label{con1}
\end{align}
Here, since we consider the matter density perturbations relevant to the LSS after the matter dominant epoch, we assumed
\begin{align}
\kappa\ ^{\mathrm{lss}}\dot{\Phi}= \kappa(\ ^{\mathrm{ad}}\dot{\Phi} + \epsilon \ ^{\mathrm{iso}}\dot{\Phi})\simeq 0.
\label{eq:dotPhi=0}
\end{align}

Next, we substitute  Eqs.~(\ref{notationa})--(\ref{notationd}) 
into the Euler equation Eq.~(\ref{eul}), which gives the zeroth order piece [$\mathcal{O}(\epsilon^0)$] to determine the adiabatic mode as
\begin{align}
&(\dot{\rho_0}(1+\kappa\ ^{\mathrm{ad}}\delta) + \rho_0\kappa \ ^{\mathrm{ad}}\dot{\delta})\kappa\ ^{\mathrm{ad}}v^{i} + \rho_0(1+\kappa\ ^{\mathrm{ad}}\delta)\kappa\ ^{\mathrm{ad}}\dot{v}^{i}+ 4H\rho_0(1+\kappa\ ^{\mathrm{ad}}\delta)\kappa\ ^{\mathrm{ad}}v^{i}\nonumber\\
&\quad + \frac{1}{a}\partial_j\left[(\rho_0(1+\kappa\ ^{\mathrm{ad}}\delta)\kappa \ ^{\mathrm{ad}}v^{i}\kappa \ ^{\mathrm{ad}}v^{j}\right] + \frac{1}{a}(\rho_0(1+\kappa\ ^{\mathrm{ad}}\delta)\partial_i\left[\kappa\ ^{\mathrm{ad}}\Psi\right] + 4\kappa\ ^{\mathrm{ad}}\dot{\Phi} \rho_0(1+\kappa\ ^{\mathrm{ad}}\delta)\kappa \ ^{\mathrm{ad}}v^{i}=0.
\label{eq:euler-0th}
\end{align}
Using Eq.~(\ref{eq:conserve}), Eq.~(\ref{eq:euler-0th}) can be simplified as
\begin{align}
&\kappa\ ^{\mathrm{ad}}\dot{v}^{i}+ H\kappa\ ^{\mathrm{ad}}v^{i} +  \frac{1}{a}\kappa\ ^{\mathrm{ad}}v^{j}\partial_j\left[\kappa\ ^{\mathrm{ad}}v^{i} \right] + \frac{1}{a}\partial_i\left[\kappa\ ^{\mathrm{ad}}\Psi\right] + \kappa\ ^{\mathrm{ad}}\dot{\Phi}  \kappa\ ^{\mathrm{ad}}v^{i} = 0.
\label{eq:euler-0th-simple}
\end{align}
The first order of $\epsilon$ in the Euler equation yields
\begin{align}
\nonumber
&\ ^{\mathrm{sh}}\dot{v}^i + \kappa\ ^{\mathrm{iso}}\dot{v}^{i} + H(\ ^{\mathrm{sh}}v^i + \kappa\ ^{\mathrm{iso}}v^{i}) +\frac{1}{a}(\ ^{\mathrm{sh}}v^j + \kappa\ ^{\mathrm{iso}}v^{j}) \partial_j\left[\kappa\ ^{\mathrm{ad}}v^{i}\right] + \frac{1}{a} \kappa\ ^{\mathrm{ad}}v^{j}\partial_j\left[\ ^{\mathrm{sh}}v^i + \kappa\ ^{\mathrm{iso}}v^{i} \right] \\ 
&\quad +\frac{1}{a}\partial_i\left[\ ^{\mathrm{sh}}\Psi + \kappa\ ^{\mathrm{iso}}\Psi\right] + \kappa\ ^{\mathrm{ad}}\dot{\Phi} (\ ^{\mathrm{sh}}v^i + \kappa\ ^{\mathrm{iso}}v^{i})+ (\ ^{\mathrm{sh}}\dot{\Phi} + \kappa\ ^{\mathrm{iso}}\dot{\Phi}) \kappa\ ^{\mathrm{ad}}v^{i} =0,
\end{align}
which leads to the Euler equation for the sh-mode at the order of ${\cal O}(\kappa^0\epsilon^1)$,
\begin{equation}
\label{sh_eul}
\ ^{\mathrm{sh}}\dot{v}^i  + H\ ^{\mathrm{sh}}v^i  + \frac{1}{a}\partial_i\left[\ ^{\mathrm{sh}}\Psi\right]  =0,
\end{equation}
and the remaining part reads
\begin{align}
\nonumber
& \kappa\ ^{\mathrm{iso}}\dot{v}^{i} + H\kappa\ ^{\mathrm{iso}}v^{i} +\frac{1}{a}(\ ^{\mathrm{sh}}v^j + \kappa\ ^{\mathrm{iso}}v^{j}) \partial_j\left[\kappa\ ^{\mathrm{ad}}v^{i}\right] + \frac{1}{a} \kappa\ ^{\mathrm{ad}}v^{j}\partial_j\left[\ ^{\mathrm{sh}}v^i + \kappa\ ^{\mathrm{iso}}v^{i} \right] \\ 
&\quad +\frac{1}{a}\partial_i\left[ \kappa\ ^{\mathrm{iso}}\Psi\right] + \kappa\ ^{\mathrm{ad}}\dot{\Phi} (\ ^{\mathrm{sh}}v^i + \kappa\ ^{\mathrm{iso}}v^{i})+ (\ ^{\mathrm{sh}}\dot{\Phi} + \kappa\ ^{\mathrm{iso}}\dot{\Phi}) \kappa\ ^{\mathrm{ad}}v^{i} =0,
\end{align}

Using Eqs.~(\ref{sh_con}), (\ref{eq:dotPhi=0}), and (\ref{sh_eul}), we finally obtain 
the following equations from the continuity equation and 
the Euler equation: 
\begin{eqnarray}
&&\kappa\left\{\ ^{\mathrm{ad}}\dot{\delta} + \frac{1}{a}\partial_i\left[\ ^{\mathrm{ad}}v^{i}\right]\right\}+ \kappa^2\frac{1}{a}\partial_i\left[ ^{\mathrm{ad}}\delta\ ^{\mathrm{ad}}v^{i}\right] =0,
\label{scp1}
\\
&&\kappa\left\{\ ^{\mathrm{ad}}\dot{v}^{i}+ H\ ^{\mathrm{ad}}v^{i}  + \frac{1}{a}\partial_i\left[\ ^{\mathrm{ad}}\Psi\right]\right\} +  \kappa^2\frac{1}{a} 
\ ^{\mathrm{ad}}v^{j}\partial_j\left[\ ^{\mathrm{ad}}v^{i} \right]= 0,
\label{scp2}
\end{eqnarray}
at the order of $\mathcal{O}(\epsilon^0)$, and 
\begin{eqnarray}
&&\kappa\left\{\ ^{\mathrm{iso}}\dot{\delta}
+ 3\ ^{\mathrm{ad}}\delta\ ^{\mathrm{sh}}\dot{\Phi} 
+ \frac{1}{a}\partial_i\left[\ ^{\mathrm{iso}}v^{i}  + \ ^{\mathrm{ad}}\delta\ ^{\mathrm{sh}}v^i  + \ ^{\mathrm{sh}}\delta \ ^{\mathrm{ad}}v^{i}\right]\right\} +\kappa^2\left\{\frac{1}{a}\partial_i\left[\ ^{\mathrm{ad}}\delta\ ^{\mathrm{iso}}v^{i}+ \ ^{\mathrm{iso}}\delta \ ^{\mathrm{ad}}v^{i}\right]\right\}
 =0,
\label{con1}\\
&&\kappa\left\{\ ^{\mathrm{iso}}\dot{v}^{i} + H \ ^{\mathrm{iso}}v^{i} + 
\ ^{\mathrm{sh}}\dot{\Phi}\ ^{\mathrm{ad}}v^{i}
+ \frac{1}{a}\left(
\ ^{\mathrm{sh}}v^j\partial_j
\left[\ ^{\mathrm{ad}}v^{i}\right] 
+\ ^{\mathrm{ad}}v^{j}\partial_j\left[\ ^{\mathrm{sh}}v^i\right]
 + \partial_i
\left[\ ^{\mathrm{iso}}\Psi\right]\right)\right\}
\nonumber\\
&&+\kappa^2\frac{1}{a}\left\{
\ ^{\mathrm{iso}}v^{j} 
\partial_j
\left[\ ^{\mathrm{ad}}v^{i}\right]
+
\ ^{\mathrm{ad}}v^{j}\partial_j
\left[\ ^{\mathrm{iso}}v^{i} \right]\right\}=0,
\label{eul1}
\end{eqnarray}
at the order of $\mathcal{O}(\epsilon^1)$, respectively. 
Equations (\ref{scp1}) and (\ref{scp2}) are the familiar equations for the 
standard cosmological perturbations theory of the matter density, 
while Eqs.~(\ref{con1}) and (\ref{eul1}) are the basic equations to 
describe the matter density perturbations under the influence of the 
sh-mode dark energy.

\subsection{Fourier modes of the order $\mathcal{O}(\kappa^1\epsilon^1)$ perturbations}
We perform Fourier transformation for the perturbations derived in
the previous section. In the following, we focus on the equations of
the order $ \epsilon $ first. The Fourier transformations of the 
perturbations are written as 
\begin{align}
\label{fourier1}
\ ^{\mathrm{lss}}\delta(t,\bm{x}) &= \frac{1}{(2\pi)^3}\int d^3p\ ^{\mathrm{lss}}\delta(t,\bm{p})e^{i\bm{p}\cdot\bm{x}},\\
\label{fourier2}
\ ^{\mathrm{lss}}v^{i}(t,\bm{x}) &=\frac{1}{(2\pi)^3}\int d^3p\frac{-ip^i}{p^2}aH\ ^{\mathrm{lss}}\theta(t,\bm{p})e^{i\bm{p}\cdot\bm{x}},\\
\label{fourier3}
\ ^{\mathrm{iso}}\Psi(t,\bm{x}) &= \frac{1}{(2\pi)^3}\int d^3p\ ^{\mathrm{iso}}\Psi(t,\bm{p})e^{i\bm{p}\cdot\bm{x}},
\end{align}
with
\begin{align}
    \ ^{\mathrm{lss}}\theta(t,\bm{x})  \equiv \frac{\nabla_i\ ^{\mathrm{lss}}v^{i}(t,\bm{x})}{aH}.
\end{align}
Here $\bm{p}$ is the wave number vector and we define $p=|\bm p|$.
Since the ad-mode and sh-mode of $\Psi$ do not contribute to density fluctuations, we write the iso-mode only for $\Psi$.

The perturbations $\delta$, $\theta$, $\Psi$ can be expanded according to standard perturbation theory with respect to $\kappa$ as
\begin{align}
\kappa\ ^{\mathrm{lss}}\delta &= \kappa\ ^{\mathrm{lss}}\delta_{\kappa} +\mathcal{O}(\kappa^2),\\
\kappa\ ^{\mathrm{lss}}\theta &= \kappa\ ^{\mathrm{lss}}\theta_{\kappa}+\mathcal{O}(\kappa^2),\\
\kappa\ ^{\mathrm{iso}}\Psi &=\kappa\ ^{\mathrm{iso}} \Psi_{\kappa}+ \mathcal{O}(\kappa^2),
\end{align}
and we ignore terms higher than the order of $\mathcal{O}(\kappa^2)$.

We consider the Fourier transform of the continuity equation Eq.~(\ref{con1}).
By using the multipole expansion Eq.~(\ref{sh-delta-dipole}), we have
\begin{align}
\ ^{\mathrm{sh}}\rho &\simeq\rho_0\sum_{n=1}^3\delta_1^{(n)}T_i^{(n)}(t)x^i,
\end{align}
with which, from Eq.~(\ref{con1}),  we obtain the continuity equation of $\mathcal{O}(\kappa \epsilon)$  as
\begin{align}
&\frac{1}{(2\pi)^3}\int d^3k_1\ ^{\mathrm{iso}}\dot{\delta}_{\kappa}(t,\bm{k}_1)e^{i\bm{k}_1\cdot\bm{x}} +  H\frac{1}{(2\pi)^3}\int d^3k_2\ ^{\mathrm{iso}}\theta_{\kappa}(t,\bm{k}_2)e^{i\bm{k}_2\cdot\bm{x}}+\sum_{n=1}^3V^{(n)}_1(t)T_i^{(n)}\frac{1}{a}\frac{1}{(2\pi)^3}\int d^3k_1(ik_1^i)\ ^{\mathrm{ad}}\delta_{\kappa}(t,\bm{k}_1)e^{i\bm{k}_1\cdot\bm{x}}\nonumber\\
&\quad +\frac{1}{a}\sum_{n=1}^3\delta_1^{(n)}(t)T_i^{(n)}\frac{1}{(2\pi)^3}\int d^3k_1\frac{-ik_1^i}{k_1^2}aH\ ^{\mathrm{ad}}\theta_{\kappa}(t,\bm{k}_1)e^{i\bm{k}_1\cdot\bm{x}} +iH\sum_{n=1}^3\delta_1^{(n)}(t)T_j^{(n)} \frac{1}{(2\pi)^3}\int d^3k_1\frac{\partial}{\partial k_1^j}\left[\ ^{\mathrm{ad}}\theta_{\kappa}(t,\bm{k}_1)\right]e^{i\bm{k}_1\cdot\bm{x}}\nonumber\\
&\quad +3\sum_{n=1}^3\dot{\Phi}_1^{(n)}(t)T_i^{(n)} \frac{1}{(2\pi)^3}\int d^3k_1\frac{\partial}{\partial k^i_1}\left[\ ^{\mathrm{ad}}\delta_{\kappa}(t,\bm{k}_1)\right]ie^{i\bm{k}_1\cdot\bm{x}}
=0.
\label{eq:continuity}
\end{align}
Hereafter, we omit $\kappa$ and  $\epsilon$ in the expressions whose order of perturbation are understood as $\mathcal{O}(\kappa \epsilon)$. 

By applying the Fourier transformations in each term of Eq.~(\ref{eq:continuity}), we rewrite it as	
\begin{align}
&\frac{1}{H}\ ^{\mathrm{iso}}\dot{\delta}_{\kappa}(t,\bm{p}) +  \ ^{\mathrm{iso}}\theta_{\kappa}(t,\bm{p}) + \sum_{n=1}^3V^{(n)}_1(t)T_i^{(n)}\frac{1}{aH}(ip^i)\ ^{\mathrm{ad}}\delta_{\kappa}(t,\bm{p}) 
 +\sum_{n=1}^3\delta_1^{(n)}(t)T_i^{(n)}\frac{-ip^i}{p^2}\ ^{\mathrm{ad}}\theta_{\kappa}(t,\bm{p})\nonumber\\ 
&\quad+i\sum_{n=1}^3\delta_1^{(n)}(t)T_j^{(n)}\frac{\partial}{\partial p^j}\left[\ ^{\mathrm{ad}}\theta_{\kappa}(t,\bm{p})\right]
+3i\frac{1}{H}\sum_{n=1}^3\dot{\Phi}_1^{(n)}(t)T_i^{(n)}\frac{\partial}{\partial p^i}\left[\ ^{\mathrm{ad}}\delta_{\kappa}(t,\bm{p})\right]
=0.
\label{cont-purt1}
\end{align}

Similarly, we obtain the Euler equation of $\mathcal{O}(\kappa\epsilon)$ after the Fourier expansions to Eq.~(\ref{eul1}),
\begin{align}
&\frac{1}{(2\pi)^3}\int d^3k_1\frac{-ik_1^i}{k_1^2}(\ddot{a}\ ^{\mathrm{iso}}\theta_{\kappa}(t,\bm{k}_1) + \dot{a}\ ^{\mathrm{iso}}\dot{\theta}_{\kappa}(t,\bm{k}_1))e^{i\bm{k}_1\cdot\bm{x}} + \frac{1}{(2\pi)^3}\int d^3k_1\frac{-ik_1^i}{k_1^2}aH^2\ ^{\mathrm{iso}}\theta_{\kappa}(t,\bm{k}_1)e^{i\bm{k}_1\cdot\bm{x}} \nonumber \\
&\quad + \frac{1}{a}\frac{1}{(2\pi)^3}\int d^3k_1(ik_1^i)\ ^{\mathrm{iso}}\Psi_{\kappa}(t,\bm{k}_1)e^{i\bm{k}_1\cdot\bm{x}} + \frac{1}{a}{\ ^{\mathrm{sh}}v^j}\frac{1}{(2\pi)^3}\int d^3k_1\frac{(k_1^i)(k_1^j)}{k_1^2}aH\ ^{\mathrm{ad}}\theta_{\kappa}(t,\bm{k}_1)e^{i\bm{k}_1\cdot\bm{x}} \nonumber \\
&\quad + \frac{1}{a} \frac{1}{(2\pi)^3}\int d^3k_1\frac{-ik_1^j}{k_1^2}aH\ ^{\mathrm{ad}}\theta_{\kappa}(t,\bm{k}_1)e^{i\bm{k}_1\cdot\bm{x}}\partial_j\left[\ ^{\mathrm{sh}}v^i\right]  +\ ^{\mathrm{sh}}\dot{\Phi}\frac{1}{(2\pi)^3}\int d^3k_1\frac{-ik_1^i}{k_1^2}aH\ ^{\mathrm{ad}}\theta_{\kappa}(t,\bm{k}_1)e^{i\bm{k}_1\cdot\bm{x}}=0.
\end{align}
Note that $\partial_j\left[\ ^{\mathrm{sh}}v^i\right]=0$ if we recall that $\ ^{\mathrm{sh}}v^i$ is a spatially uniform field independent of the coordinate $x^{i}$ [see Eq.~(\ref{sh-v-dipole})].
Hence, substituting the multipole expansions Eqs.~(\ref{sh-v-dipole}) and (\ref{sh-Phi-dipole}), we obtain the following Euler equation of $\mathcal{O}(\kappa\epsilon)$:
\begin{align}
&\frac{1}{H}\ ^{\mathrm{{iso}}}\dot{\theta}_{\kappa}(t,\bm{p}) + (\frac{\dot{H}}{H^2} + 2)\ ^{\mathrm{iso}}\theta_{\kappa}(t,\bm{p})   {-\frac{p^2}{a^2H^2}\ ^{\mathrm{iso}}\Psi_{\kappa}(t,\bm{p})} \nonumber \\
&+ \frac{i}{aH}\sum_{n=1}^3V^{(n)}_1(t)T_j^{(n)} p^j\ ^{\mathrm{ad}}\theta_{\kappa}(t,\bm{p}) +\frac{1}{H}\sum_{n=1}^3\dot{\Phi}_1^{(n)}(t)ip^iT_{{j}}^{(n)} \frac{\partial}{\partial p^{{j}}}
\left[\frac{p^i}{p^2}\ ^{\mathrm{ad}}\theta_\kappa(t,\bm{p})\right] 
=0.
\label{eul-purt1}
\end{align}

\subsection{\label{subsec:formulation}The power spectrum of structure formation}

From the cosmological Poisson equation (for scales with wave number $k\gg aH$), we have
\begin{align}
\label{poisson}
{}^{\mathrm{lss}}\Psi = -\frac{4\pi G a^2 \rho_0}{p^2}{}^{\mathrm{lss}}\delta, 
\end{align}
which is also applicable for both the ad-mode ${}^{\mathrm{ad}}\delta$ and the iso-mode ${}^{\mathrm{iso}}\delta$ as they sum up to be the lss-mode by definition, that ${}^{\mathrm{lss}}\delta\equiv{}^{\mathrm{ad}}\delta+{}^{\mathrm{iso}}\delta$.
Furthermore, for the familiar ad-mode ${}^{\mathrm{ad}}\delta$, we may apply standard cosmological perturbation theory up to the linear order $\mathcal{O}(\kappa)$; hence, the equation reads
\begin{align}
&^{\mathrm{ad}}\ddot{\delta}_{\kappa}(t,\bm{p}) + 2H  ^{\mathrm{ad}}\dot{\delta}_{\kappa}(t,\bm{p}) - 4\pi G \rho_0\ ^{\mathrm{ad}}\delta_{\kappa} =0,
\label{eom-admode}
\end{align}
whose standard solutions are written as
\begin{align}
\label{deltaad}
\ ^{\mathrm{ad}}\delta_{\kappa}&= D_1(t)\delta_L(\bm{p}),\\
\label{thetaad}
\ ^{\mathrm{ad}}\theta_{\kappa} &= -f(t)D_1(t)\delta_L(\bm{p}),\\
\label{growthrate}
f(t) &= \frac{\mathrm{d}(\ln{D_1})}{\mathrm{d}(\ln{a})}=\frac{1}{H}\frac{\dot{D}_1}{D_1}.
\end{align}
Here, $D_1(t)\propto {}^{\rm ad}\delta(t)$ is the growth factor of the growing mode, $\delta_L(\bm{p})$ a constant depending on the Gaussian distribution of the initial density fluctuations, and $f(t)$ the linear growth rate. 

We can use Eq.~(\ref{cont-purt1}) to obtain the $\ ^{\mathrm{iso}}\theta_\kappa$ expressed with $\ ^{\rm iso}\delta_\kappa$ and the ad-mode $\ ^{\rm ad}\delta_\kappa$, $\ ^{\rm ad}\theta_\kappa$. 
Subsequently, inserting the  obtained $\ ^{\mathrm{iso}}\theta_\kappa$ into Eq.~(\ref{eul-purt1}) and applying Eqs.~(\ref{poisson})--(\ref{growthrate}), we reach the equation that $\ ^{\mathrm{iso}}\delta_{\kappa}$ 
follows, which can be written as
\begin{align}
\label{ODEdelta}
&^{\mathrm{iso}}\ddot{\delta}_{\kappa}(t,\bm{p}) + 2H  ^{\mathrm{iso}}\dot{\delta}_{\kappa}(t,\bm{p}) - 4\pi G \rho_0\ ^{\mathrm{iso}}\delta_{\kappa}  \equiv S_{\delta}(t,\bm{p}).
\end{align}
Here the source term $S_{\delta}(t,\bm{p})$ can be written explicitly to the order $\mathcal{O}(\kappa\epsilon)$ that we are interested in, as
\begin{align}
\label{sourceterm}
S_{\delta}(t,\bm{p})&\equiv      - \sum_{n=1}^3V^{(n)}_1(t)T_i^{(n)}\frac{1}{a}(ip^i)HD_1(t)\delta_L(\bm{p})(2f(t) + 1) - \frac{1}{a}\sum_{n=1}^3\dot{V}^{(n)}_1T_i^{(n)}(ip^i)D_1(t)\delta_L(\bm{p}) +\mathcal{O}\bigl((H_0/p)^0\bigr),
\end{align}
\noindent
whose generalized form in higher order of perturbations for the inhomogeneities is noted in Appendix~\ref{AppendixC} as Eq.~(\ref{sourcetermfull}). 
Specially, we note here that the usage of subscripts ``$1$'' and ``$2$'' of the growth mode $D_1$ and decay mode $D_2$ follows the convention, which should not be confused with all other cases where subscript ``$1$'' combined with a superscript ``$(n)$'' denote the dipole of the perturbations in the following parts, for example, most importantly in $V_1^{(n)}$. 
We also note that ${\cal O}\bigl((H_0/p)^0\bigr)$ denotes the higher order terms of 
the expansion $(H_0/p)^n$ with $n\geq 0$.

The source term $S_{\delta}(t,\bm{p})$ for the solution of the iso-mode arises from the existence of the sh-mode; physically, it represents the inhomogeneity in matter distribution induced by the existence of the dark energy inhomogeneity.
The solution of Eq.~(\ref{ODEdelta}) is 
\begin{align}
\ ^{\mathrm{iso}}\delta_{\kappa}(t,\bm{p}) = c_{1}(\bm{p})D_1(t) + c_2(\bm{p})D_2(t) + \int^t_0dt'\frac{D_1(t')D_2(t) -D_1(t)D_2(t')}{W\left[D_1(t'),D_2(t')\right]}S_{\delta}(t',\bm{p}),
\label{soldeltaiso}
\end{align}
where $W\left[D_1(t'),D_2(t')\right]$ is the Wronskian defined as
\begin{align}
W\left[D_1(t),D_2(t)\right] = D_1(t)\dot{D_2}(t) - \dot{D_1}(t)D_2(t),
\end{align}
and $D_2$ is the decay mode of $\ ^{\mathrm{ad}}\delta$ that vanishes in late times. 
The first term and second term of Eq.~(\ref{soldeltaiso}) are the homogeneous solutions which correspond to the well-known normal ad-mode,
the third term is a particular solution related to the source term arising from the existence of the spatial inhomogeneities of the sh-mode mentioned previously.
If the sh-mode of dark energy does not exist, then by definition of the perturbations in Sec.~\ref{sec:citeref}, no iso-mode additional to the ad-mode will arise; hence, $\ ^{\mathrm{iso}}\delta_{\kappa}(t,\bm{p})=0$. Therefore, only the particular solution part related to the source term is nontrivial for the isocurvature perturbations (iso-mode) to matter distribution to be considered in the followings. Hence, we write
\begin{align}
\ ^{\mathrm{iso}}\delta_{\kappa}(t,\bm{p}) &=\int^t_0dt'\frac{D_1(t')D_2(t) -D_1(t)D_2(t')}{W\left[D_1(t'),D_2(t')\right]}S_{\delta}(t',\bm{p}).
\end{align}
\noindent
Since $\ ^{\mathrm{iso}}\delta_{\kappa}$ are perturbations that can be comprehended as the modifications to the adiabatic mode, it is important on relatively short wavelengths with large wave numbers $p$. Therefore, if we keep terms to order  $\mathcal{O}(p)$ and neglect the assumed small quadrupole contribution, the analytical solution can be approximated as
\begin{align}
\ ^{\mathrm{iso}}\delta_{\kappa}(t,\bm{p}) \simeq & \int^t_0dt'\frac{D_1(t')D_2(t) -D_1(t)D_2(t')}{W\left[D_1(t'),D_2(t')\right]}\nonumber\\
&\times\left\{ - \sum_{n=1}^3V^{(n)}_1(t')T_i^{(n)}\frac{1}{a}(ip^i)HD_1(t')\delta_L(\bm{p})(2f(t') + 1)- \frac{1}{a}\sum_{n=1}^3\dot{V}^{(n)}_1(t')T_i^{(n)}(ip^i)D_1(t')\delta_L(\bm{p}) \right\}.
\label{deltaisot}
\end{align}
Furthermore, from Eq.~(\ref{deltaisot}), transforming the variable from $t$ to $a(t)$, we have
\begin{align}
\ ^{\mathrm{iso}}\delta_{\kappa}(a,\bm{p}) \simeq
&\sqrt{\frac{3}{4\pi}}\left(i\sum_{n=1}^3p^{n}e^{(n)}\right)\int^a_0\frac{1}{a'H(a')}da'\frac{D_1(a')D_2(a) -D_1(a)D_2(a')}{W\left[D_1(a'),D_2(a')\right]}\nonumber\\
&\times\left\{ - 3V_1(a')\frac{1}{a'}HD_1(a')\delta_L(\bm{p})(2f(a') + 1) - \frac{1}{a'}a'H\frac{\mathrm{d}V_1}{\mathrm{d}a'}D_1(a')\delta_L(\bm{p})\right\},
\label{deltaiso}
\end{align}
where we rewrote the velocity fields of the sh-mode as
\begin{eqnarray}
V^{(n)}_1(a)=e^{(n)}V_1(a),
\label{defen}
\end{eqnarray}
where ${\bm e}=(e^{(1)}, e^{(2)}, e^{(3)})$ is the unit vector. 
With the solutions for the density fluctuations, we calculate the power spectrum.

To go further, we calculate the Wronskian, whose time derivative is
\begin{align}
\dot{W}\left[D_1(t),D_2(t)\right]&= D_1(t)\ddot{D}_2(t) -\ddot{D}_1(t)D_2(t)
\label{wronskian}
\end{align}
by definition. $D_1$ and $D_2$ follow the equation of motion of the ad-mode Eq.~(\ref{eom-admode}), for example, 
\begin{align}
\ddot{D}_1 +2H\dot{D}_1 - 4\pi G\rho_0 D_1 =0.
\end{align}
Following these relation, the Wronskian can be rewritten as
\begin{align}
& D_1(t)\ddot{D}_2(t) -\ddot{D}_1(t)D_2(t)
= -2HW\left[D_1(t),D_2(t)\right].
\label{83}
\end{align}
Using Eqs.~(\ref{wronskian}) and (\ref{83}), we can obtain
\begin{align}
\frac{\dot{W}}{W} =-2\frac{\dot{a}}{a},
\end{align}
whose solution is
\begin{align}
 W = Ca^{-2},
\end{align}
where $C$ is a constant to be determined in the following.
In the limit of $a\rightarrow0$ in matter-dominant epoch, the Wronskian yields 
\begin{align}
&W\left[D_1(a'),D_2(a')\right]\simeq -\frac{5}{2}H_0\sqrt{\Omega_m}a^{-2},
\end{align}
where we have found the constant $C = -\frac{5}{2}H_0\sqrt{\Omega_m}$ by using the initial condition similar to the $\Lambda$CDM model.

Using the previous expressions of the Wronskian $W$ to compute $\ ^{\mathrm{iso}}\delta_{\kappa}(a,\bm{p})$, we obtain
\begin{align}
\ ^{\mathrm{iso}}\delta_{\kappa}(a,\bm{p}) 
&=\frac{2}{5}\sqrt{\frac{3}{4\pi}}\frac{i\sum_{n}p^{n}e^{(n)}}{ H_0\sqrt{\Omega_m}}\delta_L(\bm{p})D_2(a)\int^a_0da'\left(2D^2_1(a') f(a')V_1(a') + D^2_1(a')V_1(a') + D^2_1(a')a'\frac{\mathrm{d}V_1}{\mathrm{d}a'}\right)\nonumber \\
&\qquad -\frac{2}{5}\sqrt{\frac{3}{4\pi}}\frac{i\sum_{n}p^{n}e^{(n)}}{ H_0\sqrt{\Omega_m}}D_1(a)\delta_L(\bm{p})\int^a_0da' D_2(a')\left\{V_1(a')\left(2 f(a')D_1(a') + D_1(a')\right) + D_1(a')a'\frac{\mathrm{d}V_1}{\mathrm{d}a'}\right\}\nonumber\\
&\equiv\frac{2}{5}D_1(a)\sqrt{\frac{3}{4\pi}}\frac{i\sum_{n}p^{n}e^{(n)}}{H_0\sqrt{\Omega_m}}\delta_L(\bm{p})\left(\mathcal{I}(a)-\mathcal{J}(a)\right).
\end{align}
Here we have defined
\begin{align}
\mathcal{I}(a)&\equiv \frac{D_2(a)}{D_1(a)}\int^a_0 da'D_1(a')\mathcal{G}(a'),
\label{q1n}
\\ 
\mathcal{J}(a) &\equiv \int^a_0da' D_2(a')\mathcal{G}(a'),
\label{q2n}
\end{align}
with the growth kernel $\mathcal{G}(a)$ 
defined by
\begin{align}
\mathcal{G}(a) &\equiv D_1(a)\left\{ V_1(a)\left[2f(a)+1\right] + a\frac{\mathrm{d}V_1(a)}{\mathrm{d}a}\right\},
\label{g_kernel}
\end{align}
which arises from the coupling of the sh-mode $V_1(a)$ and ad-mode $D_1(a)$.
Therefore, the density fluctuations of the lss-mode up to $\mathcal{O}(\kappa)$ can be written as
\begin{align}
\ ^{\mathrm{lss}}\delta &= \ ^{\mathrm{ad}}\delta_{\kappa} + \epsilon \ ^{\mathrm{iso}}\delta_{\kappa} \nonumber \\
&= D_1(a)\delta_L(\bm{p}) +\epsilon\frac{2}{5}\sqrt{\frac{3}{4\pi}}D_1(a)\frac{i\sum_{n}p^{n}e^{(n)}}{H_0\sqrt{\Omega_m}}\delta_L(\bm{p})\left(\mathcal{I}(a)-\mathcal{J}(a)\right).
\end{align}

 By using the above results, we evaluate the matter power spectrum $P(a,\bm{k})$
 defined by
\begin{align}
\left<\ ^{\mathrm{lss}}\delta(a,\bm{k}_1)\ ^{\mathrm{lss}}\delta(a,\bm{k}_2)\right>
=P(a,\bm{k}_1)\delta^{(3)}_D(\bm{k}_1+\bm{k}_2),
\end{align}
where $\delta^{(3)}_D(\bm{k}_1+\bm{k}_2)$ is the Dirac delta function.
Because of the following expectation value 
\begin{eqnarray}
\left<\ ^{\mathrm{lss}}\delta(a,\bm{k}_1)\ ^{\mathrm{lss}}\delta(a,\bm{k}_2)\right> &=&D^2_1(a)\left<\delta_L(\bm{k}_1)\delta_L(\bm{k}_2)\right> 
\nonumber\\
&&- \epsilon^2\frac{3}{25\pi}\frac{\sum_{n,n'} k_1^{n} e^{(n)}k_2^{n'}e^{(n')}}{H^2_0\Omega_m}\left(\mathcal{I}(a)-\mathcal{J}(a)\right)^2D^2_1(a)\left<\delta_L(\bm{k}_1)\delta_L(\bm{k}_2)\right>,
\label{eq:powerspectrum}
\end{eqnarray}
we finally obtain
the power spectrum under the influence of the dark energy inhomogeneity as
\begin{eqnarray}
P(a,\bm{k})=P(a,k,\theta)=P_0(a,k)\left(1+ \epsilon^2 k^2\cos^2\theta {\cal R}(a)\right),
\label{Pak}
\end{eqnarray}
where we used the relation
\begin{align}
\left<\delta_L(\bm{k}_1)\delta_L(\bm{k}_2)\right> = P_m(k_1)\delta^{(3)}_D(\bm{k}_1+\bm{k}_2),
\label{eq:P_0}
\end{align}
with the matter power spectrum in a homogeneous universe $P_m(k)$, 
\begin{align}
P_0(a,k)= D_1^2(a) P_m(k)
\label{Pak_standard}.
\end{align}
Here we defined time evolution of the relative correction as
\begin{eqnarray}
{\cal R}(a) 
&=&
\frac{3}{25\pi}\frac{1}{\Omega_m H_0^2}
\left(\mathcal{I}(a)-\mathcal{J}(a)\right)^2
\label{defcalR}
\end{eqnarray}
and 
\begin{eqnarray}
{\sum_{n=1}^3k^{n}e^{(n)}= k\cos\theta},
\label{deftheta}
\end{eqnarray}
where $\theta$ is the angle between the wave number vector $\bm k$ and the direction of the superhorizon velocity.

Now we have completed the basic formulation of the LSS under
the influence of the superhorizon inhomogeneities of dark energy, which is used for the theoretical predictions in the next section. 

\section{\label{sec:level4}Numerical evaluation and Results}
Since we have obtained the formulation of the calculation, the next step is to evaluate the power spectrum $P(a,\bm{k})$ numerically. For this purpose, we need to consider the limits on the amplitudes of the perturbations in the formulations.

\subsection{Amplitude of the perturbations arising from the sh-mode}
\label{observationalconstraint}
We may obtain the maximum allowed value of $\epsilon$ from the integrated Sachs-Wolfe (ISW) effect of the CMB temperature anisotropies \cite{nan2019large,nan2021dark}.
The ISW effect on the CMB temperature anisotropies 
coming from the superhorizon dark energy can be estimated by 
\begin{align}
\frac{\Delta T}{T}({\bm \gamma}) &
\simeq 2\epsilon\int_{\eta_d}^{\eta_0} \mathrm{d}\eta\left({\partial\ ^{\rm sh}\Psi(\eta,\chi,{\bm \gamma}
)\over\partial\eta}\right)\Bigg|_{\chi=\eta_0-\eta}
  \simeq
  2\epsilon \int_{\eta_d}^{\eta_0} \mathrm{d}\eta \left(
  \sum^3_{n=1}{\partial\Psi_{1}^{(n)}(\eta)\over\partial\eta} T^{(n)}_i \chi\gamma^i
  \right) \Bigg|_{\chi=\eta_0-\eta},
\label{dtotg}
\end{align}
where $\bm \gamma$ is the unit vector of the line of sight direction, and we used that the comoving coordinate $x^i$ is written as $x^i=\chi \gamma^i$ using the radial coordinate $\chi$ and the component of $\bm \gamma$. 
It can also be confirmed that the matrices $T^{(n)}_i$ introduced in Sec.~\ref{subsec:sh-mode} are related to the real basis spherical harmonics as
\begin{align}
Y_{\ell=1}^{(n)}(\omega,\varphi) \equiv T_i^{(n)}{x^i\over \chi},
\label{eq:harmonics}
\end{align}
where we used $\bm \gamma\equiv(\sin\omega\cos\varphi,\sin\omega\sin\varphi,\cos\omega)$. Then, we can rewrite Eq.~(\ref{dtotg}) as  
\begin{align}
\frac{\Delta T}{T}(\bm \gamma) &= 2 \epsilon\sum_{n=1}^3J_{1}^{(n)}Y_1^{(n)}(\omega,\varphi),
\end{align}
where we defined
\begin{align}
J_{1}^{(n)}\equiv \int_{\eta_d}^{\eta_0}\mathrm{d}\eta(\eta_0-\eta) \left(\frac{\partial \Psi_1^{(n)}(\eta)}{\partial \eta}\right),
\end{align}
where $\eta_0$ and $\eta_d$ are the conformal time at the present epoch and the decoupling time, respectively.
Using the previous work in Ref.~\cite{nan2021dark}, the solution of $J_1^{(n)}$ can be 
obtained, as demonstrated in Table~\ref{table:paraset}, with which we can put a constraint 
on the sh-mode from the observation of the CMB as 
\begin{align}
\frac{(2\epsilon)^2}{3}\sum_{n=1}^3(J_1^{(n)})^2 \leq C^{\rm{obs}}_1,
\end{align}
where $C_1^{\rm obs}$ is the dipole component of the multipole expansion of the 
angular correlation function of the CMB temperature anisotropies, 
$C(\vartheta)=\sum_{\ell}C_\ell^{\rm obs} P_\ell(\cos\vartheta){(2\ell+1)/ 4\pi}$.

Using the Planck Legacy Archive,\footnote{Based on observations obtained with Planck (http://www.esa.int/Planck), an ESA science mission with instruments and contributions directly funded by ESA Member States, NASA, and Canada.} we have $C^{\rm{obs}}_1 \leq 6.3 \times 10^{-6}$, and obtain the maximum allowed value 
of $\epsilon$ as 
\begin{align}
\epsilon_{\rm max} \left[\sum_{n=1}^3 (J_1^{(n)})^2\right]^{1/2}\leq 2.2\times 10^{-3}. 
\label{maxCMB}
\end{align}
We use this maximum allowed value for the numerical calculation of the power spectrum in the next subsection.

\subsection{\label{subsec:P(k)}Power spectrum and $\sigma_8$}
\begin{figure}[t]
\begin{center}
\includegraphics[width=0.8\linewidth,pagebox=cropbox,clip]{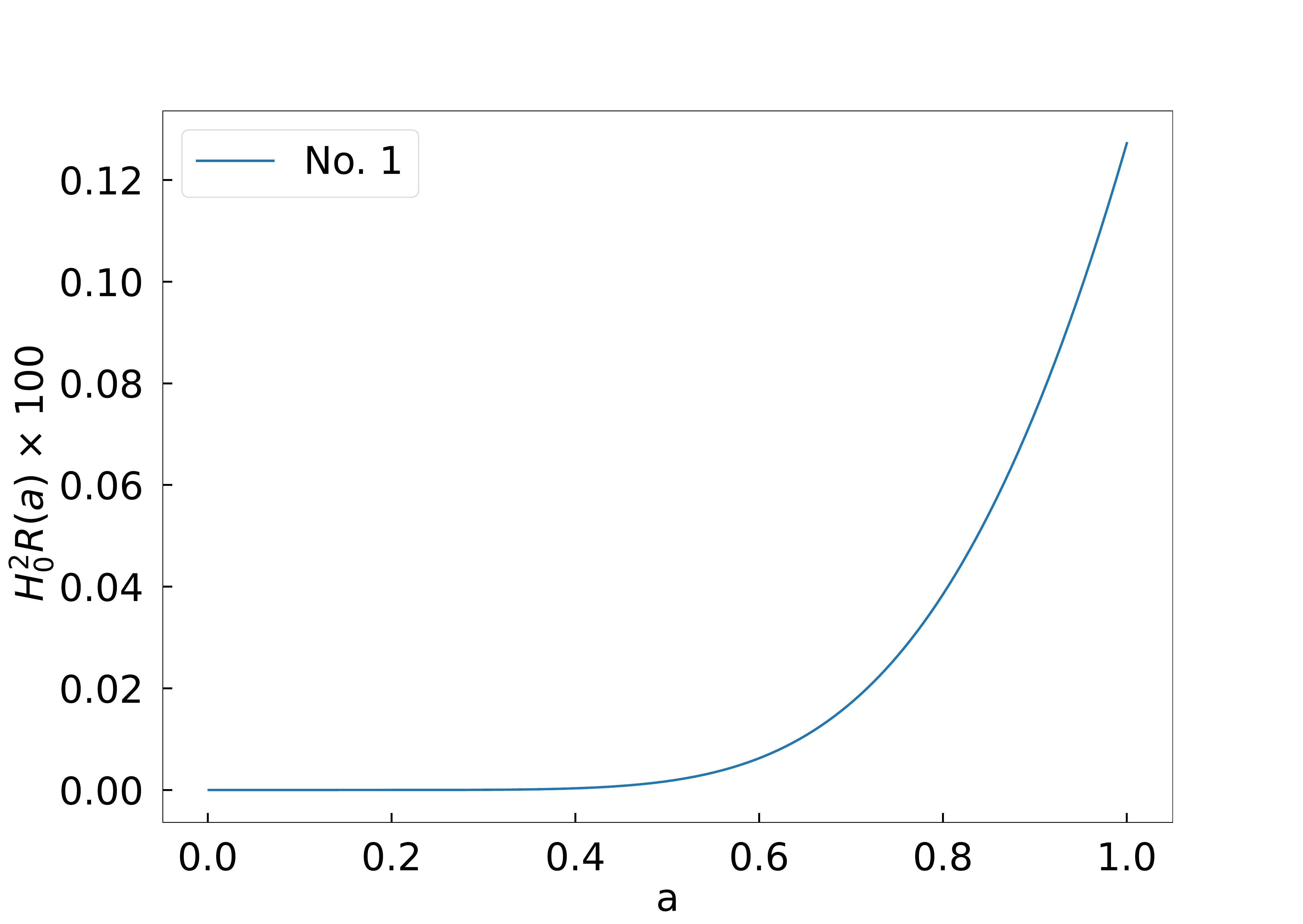}
\end{center}
\caption{This figure plots the time-evolution behavior of the modification term $\mathcal{R}(a)$ to the matter power spectrum defined in Eq.~(\ref{defcalR}). The modification to the matter power spectrum grows rapidly after $a\gtrsim0.5$ when the dark energy becomes important; hence, the sh-mode perturbation associated with the inhomogeneity of the dark energy assumed in this model begins to show its impact on structure formation. Notice that the curve and the ticks of the vertical axis are multiplied by a factor of 100.
We here adopt the model No.~1 in Table~\ref{table:paraset} as an example.
}
\label{fig:Ra}
\end{figure}

We here focus on the power spectrum described by Eq.~(\ref{Pak}). 
The relative difference of the power spectrum is given by 
\begin{eqnarray}
\xi_{\rm modif}(k,\theta,a)\equiv{P(a,k,\theta)\over P_0(a,k)}-1= \epsilon_{\rm max}^2 k^2\cos^2\theta {\cal R}(a),
\label{Pak2}
\end{eqnarray}
where ${\cal R}(a)$ is defined by Eq.~(\ref{defcalR}).
 Figure~\ref{fig:Ra} plots the time evolution of $\mathcal{R}(a)$.
We note that $\theta$ means the angle between the dipole direction and the wave number vector as defined by Eq.~(\ref{deftheta}). 
Each curve in this figure adopts the four sets of the parameters in Table I, which 
can be consistent with cosmological observations \cite{nan2021dark}.
Here we assumed the amplitude of the superhorizon mode as
the maximum value of $\epsilon$ given by Eq.~(\ref{maxCMB}).
This is equivalent to the assumption that 
the entire contribution to the dipole anisotropies in the observed CMB arises from the influence of the superhorizon mode dark energy inhomogeneities. 

\begin{table}[b]
\centering
    \begin{tabular}{c|c|c|c|c|c|c}\hline
    ~~~Model~~~&~~~$\tilde{r}$~~~&~~~$\tilde{m}$~~~& ~~~F~~~&~~~$\Omega_m$~~~&~~~$\epsilon_{max}$~~~&~~~$J^{(n)}_1$~~~\\ \hline \hline
    No.~1  & 70 & 1/10 & 1.00 & 0.30 &  0.0117 & 0.107\\ \hline
    No.~2  & 6.3& 1/3 & 1.01 & 0.30 & 0.0117 & 0.107\\ \hline
No.~3  & 72 & 1/10 & 1.00 & 0.28 & 0.0108 & 0.116\\ \hline
    No.~4  & 68 & 1/10 & 1.00 & 0.32 & 0.0127 & 0.0985\\ \hline
  \end{tabular}
  \caption{The parameters of the models applied for calculation in this work.}
  \label{table:paraset}
\end{table} 

\begin{figure}
\begin{center}
\includegraphics[width=0.7\linewidth,pagebox=cropbox,clip]{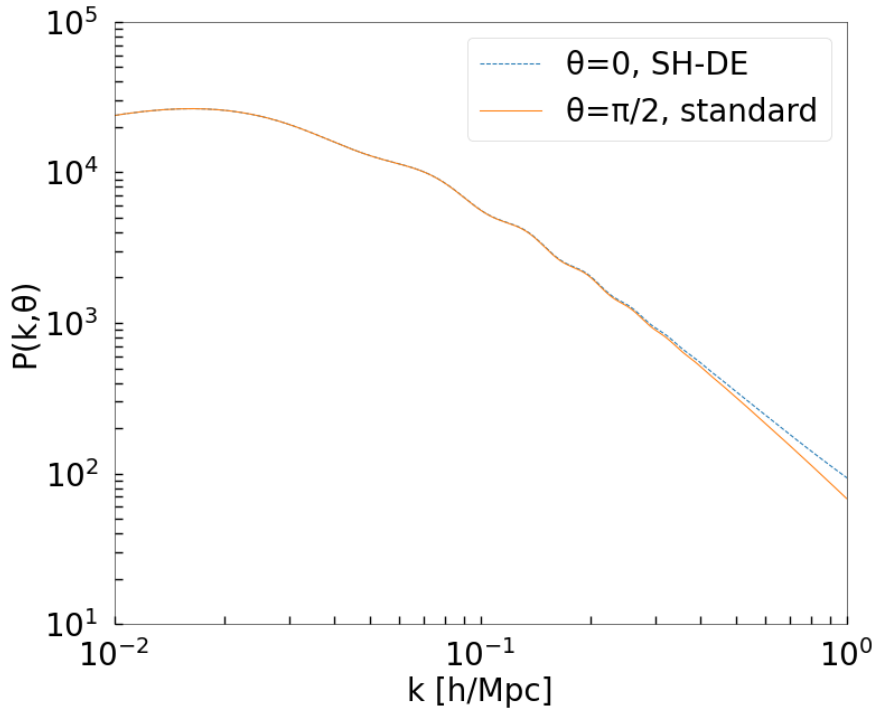}
\end{center}
\caption{This figure shows the modified matter power spectrum $P(a=1, \bm k)$ in Eq.~(\ref{Pak}) for $\theta=0$ compared with $\theta=\pi/2$, which is evaluated at the present epoch $a=1$. Here, the curve with $\theta=\pi/2$($\cos\theta=0)$ can also be understood as the linear power spectrum of the standard model without modification. Typically, the theoretical prediction of the modification scales as $\propto k^2$ and becomes powerful on small scales with large $k$, although we confine our discussion on the possible observational imprints of the model modification to the LSS power spectrum up to $k\sim 0.2{\rm \, h \, Mpc}^{-1}$ due to the usage of linear perturbation theory in the formulation (see also Fig.~\ref{fig:xik}), where the estimated modification is supposed small and consistent with existent observational results. In this figure, we adopted the model No.~1 in Table~\ref{table:paraset}.}
\label{fig:power}
\end{figure}

\begin{figure}
\begin{center}
\includegraphics[width=0.7\linewidth,pagebox=cropbox,clip]{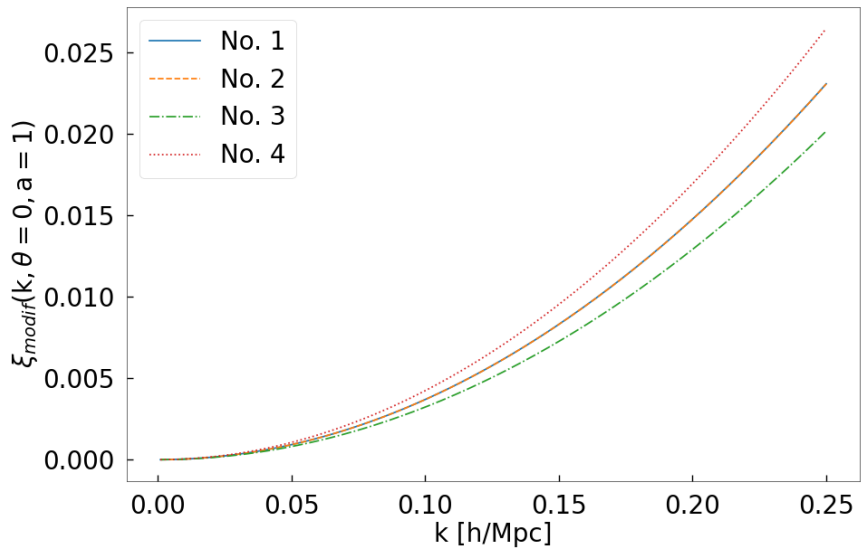}
\end{center}
\caption{Relative modification term to the power spectrum, $\xi_{\rm modif}(k, \theta,a)$ in Eq.~(\ref{Pak2}) as a function of $k$, which arises from the iso-mode.
In this figure, we fixed $\theta=0$ and $a=1$ and adopted the maximum allowed value for the initial amplitude of the sh-mode inferred from the ISW effect, Eq.~(\ref{maxCMB}). Here, we use the parameter sets shown in Table~\ref{table:paraset}. As expressed in Eq.~(\ref{Pak2}) and addressed in the caption of Fig.~\ref{fig:power}, $\xi_{\rm modif}(k, \theta,a)\propto k^2$, but the validity of the prediction may be restricted by the usage of linear perturbation theory up to $k\sim0.2{\rm \, h \, Mpc}^{-1}$. However, a possible indication of the modification to the power spectrum predicted by the model together with consideration of the quasi-linear and nonlinear regime would be interesting as a future exploration.}
\label{fig:xik}
\end{figure}

By applying the transfer function in radiation-matter domination transition and the linear matter power spectrum from the Cosmic Linear Anisotropy Solving System (CLASS)~\cite{Blas_2011} for $P_0(a, k)$ in Eq.~(\ref{Pak}), Fig.~\ref{fig:power} demonstrates the modified power spectrum predicted. The dashed curve $(\theta=0)$ and the solid curve $(\theta=\pi/2)$
correspond to $P(a=1,k,\theta=0)$ and $P(a=1,k,\theta=\pi/2)=P_0(a=1,k)$, respectively.
The quasi-nonlinear effect of the density perturbations on the power spectrum becomes
influential for $k\simgt 0.2 h{\rm Mpc}^{-1}$, which is ignored in our computation.
Because our theoretical model
relies on the linear theory of the density perturbation, then our 
theoretical predictions for the quasi-nonlinear regions $k\simgt 0.2 {\rm \, h \, Mpc}^{-1}$ 
should be carefully understood.
From the aspect of the cosmological perturbation theory, we will keep in mind that the scope of validity of $\mathcal{O}(\kappa)$ linear expansion used for $\mathcal{O}(\kappa\epsilon)$ modification to power spectrum in Fig.~\ref{fig:xik} is roughly up to $k \sim 0.2{\rm \, h \, Mpc}^{-1}$.

Figure~\ref{fig:xik} plots the relative correction to the power spectrum,
Eq.~(\ref{Pak2}), for $\theta=0$ at the present epoch $a=1$ with adopting the maximum allowed value, Eq.~(\ref{maxCMB}). This figure shows that the matter power spectrum at the present epoch may be modified by $\sim 0.4\%$ $(1\%)$ at the wave number $k\sim 0.1\rm{\, h \, Mpc}^{-1}$
$(0.15\rm{\, h \, Mpc}^{-1})$ due to the existence of dipole anisotropies in dark energy at $a = 1$.
Each curve in Fig.~\ref{fig:xik} adopts the same sets of the parameters as those of Fig.~\ref{fig:Ra}, 
denoted in Table I, where $\tilde{r},\tilde{m}$ are the parameters characterizing the superhorizon dark energy model defined as $\tilde{r} \equiv \frac{1}{6}({\overline{\phi}_0}/{M_{pl}})^2$, $\tilde{m} \equiv m/ H_0$ with $\phi_0 \equiv\overline{\phi}_0\tilde{\phi}_0$, and $F$ is a
constant used in the numerical computation (see also Appendix \ref{AppendixB}).

From Figs.~\ref{fig:Ra} and~\ref{fig:xik}, we find that power spectrum $P(a,k,\theta)$ depends
on $\Omega_m$; however, it does not much depend on the parameters $\tilde{r}$ and $\tilde{m}$. 
This is because we fixed the amplitude of the inhomogeneities of the superhorizon mode 
dark energy $\epsilon_{\rm max}$ from Eq.~(\ref{maxCMB}). 
The parameters $\tilde{r}$ and $\tilde{m}$ change the dynamics of the dark energy, 
as demonstrated in Ref.~\cite{nan2021dark}, but the predictions on the observational
quantities only depend on the amplitude of the inhomogeneities of the superhorizon mode 
dark energy $\epsilon_{\rm max}$. Therefore, the modification to the matter power spectrum
depends on $\Omega_m$ but not on $\tilde{r}$ and $\tilde{m}$.

\begin{figure}
\begin{center}
\includegraphics[width=0.7\linewidth,pagebox=cropbox,clip]{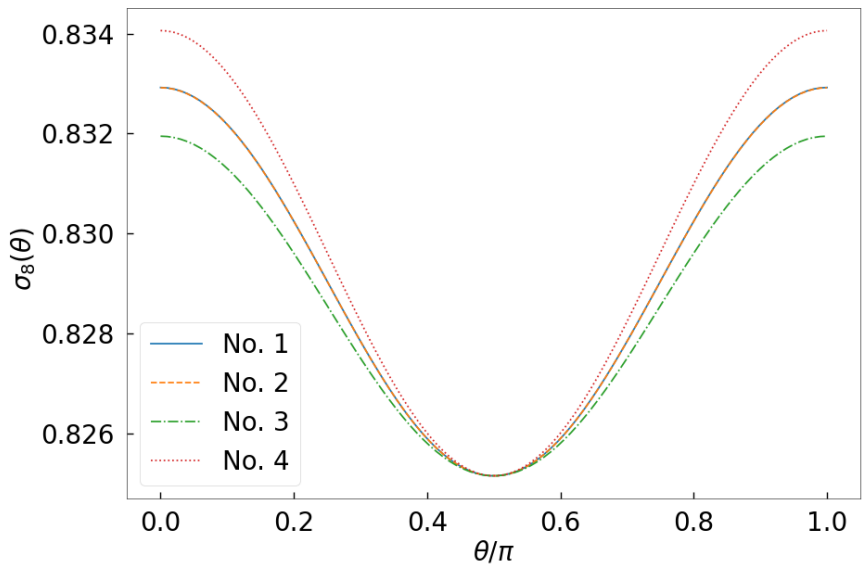}
\end{center}
\caption{The direction-dependent ${}^{\rm sh}\sigma_8(\theta)$ predicted by our model in Eq.~(\ref{sigma8}). The $\theta$ angle is related to the dipole direction of the dark energy inhomogeneities noted as the sh-mode. This figure demonstrates that, in an anisotropic way, the clumpiness of the LSS matter power spectrum on scale of $8\rm{\,h^{-1}\, Mpc}$ may be enhanced by the iso-mode perturbation to the matter introduced in this work, which is sourced by the sh-mode perturbation.
}
\label{fig:sigma8}
\end{figure}

Next, we consider the effect of the superhorizon mode dark energy on $\sigma_8$, which 
is often used as a quantity to characterize the amplitude of the matter power spectrum 
weighing the clustering of matter within radius $R$ denoted by $\sigma_R$ 
at the scale of $R=8 \rm{\,h^{-1}\,Mpc}$. 
As a forecast of order estimation, this can be evaluated roughly as $\sigma_8^2 \sim P(k)k^3|_{k =0.1{\rm \,h\, Mpc}^{-1}}$.
From this naive estimation, the correction on $P(k)$ due to the existence of the iso-mode of LSS is about $0.4\%$, and the modification to $\sigma_8$ is forecasted to be about $0.16\%$. 
However, we have obtained the numerical solutions for the sh-mode and iso-mode, and calculate $\sigma_8$ with numerical evaluations more carefully in the following. We define $\sigma_R$ in our mode including the 
inhomogeneities of the superhorizon mode dark energy by 
\begin{align}
&\ ^{\rm sh}\sigma^2_R(a,\theta) = \frac{1}{2 \pi^2}\int_0^{\infty}\mathrm{d}k k^2 P(a, k,\theta)W^2(kR),
\label{sigma8}
\end{align}
with the window function $W(kR)$ 
\begin{align}
&W(kR) = \frac{3\left( \sin{kR}-kR\cos{kR}\right)}{(kR)^3}.
\label{top-hat}
\end{align}
and Eq.~(\ref{Pak}). Figure~\ref{fig:sigma8} plots $ {}^{\rm sh}\sigma_{8}(a=1,\theta)$ as a function of $\theta$, indicating the possible anisotropic imprints of the model modification.

The maximum result for ${}^{\rm sh}\sigma_8(\theta)$ at $\cos\theta=1$ with the model No.~1 or No.~2 will be ${}^{\rm sh}\sigma_{8}(\theta=0) =0.833$, while the standard part neglecting the effects from growth 
induced by the coupling of the iso-mode and ad-mode is ${}^{\rm sh}\sigma_8(\theta=\pi/2) =0.825$. 
We conclude that the maximum correction from the iso-mode induced by superhorizon dark energy is roughly 
$1.0\%$.

\subsection{Discussion}

In Sec.~\ref{sec:level4} we presented the numerical results for the matter power spectrum 
and the cosmological parameter $\sigma_8$, 
following the theoretical model with the superhorizon dark energy inhomogeneities. 
In this section, we compare these results with the observations and discuss possible implications for further verification and application.

For the matter power spectrum $P(k)$, there have been measurements from various observations by, e.g., SDSS and DES. 
From the numerical results in Sec.~\ref{sec:level4} 
the matter power spectrum might be modified in a manner 
depending on the direction of the iso-mode by the order of $1\%$ at 
$k\sim0.15{\rm\,h\,Mpc}^{-1}$, as shown in Figure~\ref{fig:xik},
which results in the directional dependence of the $\sigma_8$ parameter
due to the superhorizon dark energy inhomogeneities. 
The numerical results in Sec.~\ref{sec:level4} show that $\sigma_8$ may
change depending on the direction by $1\%$ at the present epoch.  

On the other hand, as for recent analyses with observational measurements,  $\sigma_8$ is analyzed as $\sigma_8 = 0.829 \pm 0.015$ with Planck~\cite{Pandey_2020}, $\sigma_8 = 0.821\pm0.023$ with WMAP~\cite{Hinshaw_2013}, and $\sigma_8 = 0.785\pm0.044$ with SDSS~\cite{More_2015}.
As a comparison, the $S_8\equiv\sigma_8(\Omega_m/0.3)^{1/2}$ parameter was constrained as $S_8=0.780^{+0.030}_{-0.022}$ for flat $\Lambda$CDM with cosmic shear data analysis from the Subaru Hyper Suprime-Cam (HSC)~\cite{10.1093/pasj/psz010}, where an $S_8$ tension arises when compared with the Planck results, which is followed by the Kilo-Degree Survey (KiDS-1000) results~\cite{refId0}. However, the authors in Ref.~\cite{10.1093/mnras/stab1613} argued that it may still be premature to claim a firm tension in the $S_8$ parameter by use of the combination of $f \sigma_8$ from measurements of the redshift-space distortions.
Taking the current status of debates and systematic errors of these relevant results into account, we conclude that our model prediction is not contradictory to current observations.
This modification to $\sigma_8$ due to the iso-mode is also considered to be consistent with a previous constraint on the $\sigma_8$ modification~\cite{Hirata_2009}.
The correction indicated by the large-scale inhomogeneities from our model are within the range of allowed error, and the model may be worth future tests/analyses with observational data.

As for the $\sigma_8$ tension, the additional effect given by the coupling of the iso-mode with the ad-mode introduced by the superhorizon dark energy works in a way to boost the value of $\sigma_8$ inferred from LSS.
On the other hand, Ref.~\cite{lambiase2019testing} showed that dynamical dark energy (DDE) could ease the $\sigma_8$ tension indicated by the CMB and LSS observations compared with $\Lambda$CDM model. 
The superhorizon dark energy model adopted in the present paper is a kind of thawing quintessential DDE consistent with the Chevallier-Polarski-Linder parametrization for its equation of state~\cite{2001IJMPD..10..213C,PhysRevLett.90.091301}. 
The directional-dependent increase in $\sigma_8$, which is induced by the additional matter fluctuations (iso-mode) from DE inhomogeneities (sh-mode) modeled, could also possibly work to ease the $\sigma_8$ tension between the CMB and LSS slightly. 
For the $H_0$ tension potentially relevant to the $\sigma_8$, it could be interesting to examine the prediction of $f\sigma_8(a)$ evolution with the iso-mode of our model against the observational measurements (e.g., Ref.~\cite{PhysRevD.98.083543}) as a future investigation.
A recent work based on DDE from a rolling scalar field showed the potential of easing $H_0$ and $\sigma_8$ in the same time by special coupling with dark matter ~\cite{Agrawal:2019dlm}, motivated by the de Sitter swampland conjecture~\cite{Vafa:2005ui,PhysRevD.81.123530,Obied:2018sgi,Garg:2018reu,OOGURI2019180}, which is also relevant to the ultralight scalar field $\phi$ for superhorizon dark energy adopted in this paper.
As a comparison with the model in this work, the authors of Ref.~\cite{PhysRevD.103.L121302} show the possibility of easing the $H_0$ tension by introducing local inhomogeneities from the coupling of a chameleon dark energy model with dark matter.

Let us refocus on the hemispherical power asymmetry of the CMB. The origin of the CMB dipole is usually explained by the Galactic motions; however, it can be an open question as discussed in Refs.~\cite{2020A&A...644A.100P,sullivan2021cmb}. 
The superhorizon dark energy model adopted in this work is a possible scenario to account for the CMB dipole by introducing an intrinsic dipole
\cite{nan2021dark}.
We have formulated for the prediction of the LSS matter power spectrum in this model and checked for its consistency with observations in this work, and the model remains to be tested by future analyses of the LSS observations.  The generalization of the formulation with the source term in  Appendix~\ref{AppendixC} could be potentially interesting to investigate if an intrinsic dipole from superhorizon perturbations is confirmed/detected.

As mentioned in Sec.~\ref{sec:level1}, several studies are suggesting large-scale anomalies. 
In addition, it has been suggested that there might be a directional dependence of the fine structure constant $\alpha$, where a nonzero-dipole-component model fits better than a uniform universe model at $4.2\sigma$ level~\cite{PhysRevLett.107.191101}.
Moreover, an isotropic nonvanishing cosmic birefringence, or in other terms, CMB polarization rotation, which is of order $\mathcal{O}(0.1)$ degree, is reported by a recent analysis on the Planck 2018 polarization data at 99.2\% C.L.~\cite{PhysRevLett.125.221301}. Concerning this effect, some previous researches have studied its possible implications on dynamical dark energy models possibly related to axionlike particles (e.g. Ref.~\cite{PhysRevD.103.043509}), which is similar to the ultralight field $\phi$ of superhorizon dark energy model adopted in this work.
Hence, it would be interesting to explore whether our model could make theoretical predictions on these observations or be tested from these observations in the future, in hope of possible clues of constraints from these aspects to improve the model prediction on the LSS matter power spectrum in turn. 
Another interesting direction is to extend the formulation of superhorizon dark energy on the matter power spectrum of the LSS beyond the linear regime to examine its prediction on smaller scales, although this is beyond the scope of the present work.
Potentially related to the discrimination of dark energy models with ours using LSS as the probe,  Ref.~\cite{2018JCAP...06..018P} investigated the effect of nonlinear clustering of matter with quintessence dark energy, following which Refs.~\cite{2020CQGra..37w5008R,2021CQGra..38s5001R} showed the comparison of matter clustering with quintessence dark energy against tachyonic dark energy in the linear and nonlinear regime.

\section{\label{sec:level5}Summary and Conclusions}

In this study, motivated by observational anomalies indicating potential anisotropies beyond the cosmological principle with $\Lambda$CDM model, we formulated for the theoretical framework
to investigate LSS under the superhorizon scale dark energy inhomogeneities. 
Based on this formulation, we determined the LSS matter power spectrum of matter in the 
inhomogeneous dark energy model \cite{nan2021dark}. This model assumes an ultralight scalar field $\phi$ with $\mathcal{O}(1)$ inhomogeneities of the field configuration and the potential energy on superhorizon scales, called the sh-mode dark energy, whose inhomogeneities are small within the present horizon. 
The sh-mode dark energy causes iso-mode perturbations in addition to the usual adiabatic perturbations for the LSS matter distribution, leading to a modification to the matter power spectrum in a way that the amplitude of the perturbations depends on the dipole direction of the dark energy inhomogeneities. 

Assuming the maximum allowed value of the sh-mode to the CMB dipole, we put a constraint on the amplitude of the sh-mode. This is used to evaluate the modification to the matter power spectrum of the LSS.
With the parameters in Table~\ref{table:paraset}, we found that the modification to the matter power spectrum caused by the sh-mode scales as $k^2$, and gives modification to the matter power spectrum as $\simeq0.5\%$ at  $k\simeq 0.1 \rm{\,h\,Mpc}^{-1}$ and $\sim1\%$ at  $k\sim 0.15 \rm{\,h\,Mpc}^{-1}$. 
The modification could be large on the smaller scales, but linear perturbation theory used for the formulation may break down at scales smaller than this scale. Also, we evaluated the possible correction to $\sigma_8$, which predicted $1\%$ enhancement in the direction $\theta=0$, compared with the 
value in the direction $\theta=\pi/2$.
Our analysis contains the effects arising from superhorizon perturbations beyond the horizon. The model may also be interesting when it is investigated with the separate universe simulation method~\cite{Wagner_MNRAS_2014,Baldauf_2016,PhysRevD.100.023516} associated with the viewpoint of the ``super-sample" mode~\cite{PhysRevD.89.083519,PhysRevD.90.103530,PhysRevD.97.063527,PhysRevD.100.103515,10.1093/mnras/staa1579}, where the effects of the fluctuations with wavelength scales beyond the survey volume/region were investigated and discussed.  
By using simulations that are compatible with our model, we expect to find more clues to understand the small-scale growth in proportion to $k^2$ in the nonlinear regime of density perturbations.

\acknowledgments

This work was supported by JSPS KAKENHI Grant Number JP20J13640 (Y.N.). The authors thank the Yukawa Institute for Theoretical Physics at Kyoto University. Discussions during the YITP workshop YITP-X-21-09 "The 10th Workshop on Observational Cosmology" were useful to complete this work. Especially, we would like to thank Masahiro Takada, Daisuke Yamauchi, and Kazuyuki Akitsu for helpful comments and discussions during the workshop.

\appendix

\section{\label{AppendixA}Einstein equations with the sh-mode}

In this section, we find the Einstein tensors and energy-momentum tensors that are necessary to obtain the equations governing the evolution of the sh-mode while referring to Ref.~\cite{nan2021dark}. The metric adopted is
\begin{align}
      g_{ij} = \left(
    \begin{array}{cc}
      -(1 + 2\Psi(t,\bm{x})) & 0 \\
      0 & a^2(t)(1 + 2 \Phi(t,\bm{x}))\delta_{ij}
    \end{array}
  \right),
  \label{metric}
\end{align}
where $\Psi ={}^{\rm{sh}}\Psi$, $\Phi ={}^{\rm sh}\Phi$ for the purpose of characterizing the sh-mode.
To find the equations that the sh-mode follow,
let us start with the Einstein equation Eq.~(\ref{einstein}), which we write out again as 
\begin{align*}
G^{\mu}_{~\nu} = 8\pi G(T^{\mu(m)}_{~\nu} + T^{\mu(\phi)}_{~\nu} ),
\end{align*}
where $T^{\mu(m)}_{~\nu}$ is the energy-momentum tensor of matter, and $T^{\mu(\phi)}_{~\nu}$ is that of the scalar field $\phi$.
With the metric in Eq.~(\ref{metric}), we calculate $G^{\mu}_{~\nu}$, $T^{\mu(m)}_{~\nu}$, $T^{\mu(\phi)}_{~\nu}$ and write the explicit expressions, respectively. 
The energy-momentum tensor for scalar field is 
\begin{align}
T_{~\nu}^{\mu(\phi)} = g^{\mu \alpha}\partial_{\alpha}\phi \partial_{\nu}\phi -g^{\mu}_{~\nu}(\frac{1}{2}g^{\alpha \beta}\partial_{\alpha}\phi \partial_{\beta}\phi + \frac{1}{2}m^2\phi^2),
\end{align}
hence, we obtain its components as
\begin{align}
T_{~0}^{0(\phi)} &= -\frac{1}{2}\dot{\phi}_0^2 -\frac{1}{2}m^2\phi_0^2 + \dot{\phi}^2_0 \ ^{\mathrm{sh}}\Psi -\dot{\phi}_0\ ^{\mathrm{sh}}\dot{\Phi}-m^2\phi_0\ ^{\mathrm{sh}}\Phi,\\
T_{~i}^{0(\phi)} &= -\dot{\phi}_0\ \partial_i\ ^{\mathrm{sh}}\phi,\\
T_{~j}^{i(\phi)}&= \left(\frac{1}{2}\dot{\phi}_0^2 - \frac{1}{2}m^2\phi_0^2 - \dot{\phi}_0^2\ ^{\mathrm{sh}}\Psi + \dot{\phi}_0\ ^{\mathrm{sh}}\dot{\Phi}- m^2\phi_0\ ^{\mathrm{sh}}\Phi\right)\delta^i_{~j}\  .
\end{align}
On the other hand, for matter, $T^{\mu(m)}_{~\nu}$ is also well known, 
\begin{align}
T_{~0}^{0(m)} &=-\rho_0 - \rho_0\ ^{\mathrm{sh}}\delta ,\\
 T_{~i}^{0(m)} &= a\rho_0\ ^{\mathrm{sh}}v^i,\\
 T_{~j}^{i(m)}& 
 =0.
\end{align}
Here in the $T_{~j}^{i}$, the pressure is zero for the matter,  and the anisotropic stress $\Pi^i_{~j}$ is set to zero. Then, the components of the Einstein tensor are
\begin{align}
G^0_{~0} =& -3H^2 -6H\ ^{\mathrm{sh}}\dot{\Phi}+ 6H^2\ ^{\mathrm{sh}}\Psi + 2\frac{\Delta \ ^{\mathrm{sh}}\Phi}{a^2},\\
G^0_{~i} =& 2\partial_i\ ^{\mathrm{sh}}\dot{\Phi} -2H\partial_i\ ^{\mathrm{sh}}\Psi ,\\
G^i_{~j} =&(-2\frac{\ddot{a}}{a}-H^2)\delta^i_{~j} 
+ \frac{2}{a^2}\left[(2a\ddot{a} + \dot{a}^2)\ ^{\mathrm{sh}}\Psi + a\dot{a}\ ^{\mathrm{sh}}\dot{\Psi} -a^2\ ^{\mathrm{sh}}\ddot{\Phi} -3a\dot{a}\ ^{\mathrm{sh}}\dot{\Phi} + \frac{1}{2}\partial_i\partial_j(\Delta \ ^{\mathrm{sh}}\Psi + \Delta \ ^{\mathrm{sh}}\Phi)\right]\delta^i_{~j}\nonumber\\
&- \frac{1}{a^2}\partial_i\partial_j(\ ^{\mathrm{sh}}\Psi + \ ^{\mathrm{sh}}\Phi).
\end{align}
By substituting the above expressions into the Einstein equation Eq.~(\ref{einstein}), we obtain the background equations Eqs.~(\ref{background3})--(\ref{background4}) and sh-mode equations Eqs.~(\ref{sh5})--(\ref{sh2}). Other equations come from the equations of motion (\ref{con})--(\ref{klein-gordon}) (see also Ref.~\cite{nan2021dark}).

\newpage
\section{\label{AppendixC} The source term of the iso-mode with the quadrupole component up to $\mathcal{O}(\kappa\epsilon^2)$}
When the quadrupole component is included, instead of Eq.(\ref{multipole}), the superhorizon mode inhomogeneities are written as
\begin{align}
\epsilon ^{\mathrm{sh}}\phi(t,\bm{x}) &= \epsilon\sum_{n=1}^3\phi^{(n)}_1(t)T_i^{(n)}x^i 
+ \epsilon^2\sum_{n=1}^5 \phi^{(n)}_2(t)T^{(n)}_{ij}x^jx^i 
+\mathcal{O}(\epsilon^3 x^3),
\end{align}
 where $T_{ij}^{(n)}$ is identical to the definition of $P_{ij}^{(n)}$ in Appendix~A of Ref.~\cite{nan2021dark}.
 The order of the quadrupole component in the sense of perturbation order is ${\cal O} (\kappa\epsilon^2)$.
In this case, Eq.~(\ref{ODEdelta}) holds by replacing the source term  
Eq.~(\ref{sourceterm}) with
\begin{align}
S_{\delta}(t,\bm{p})&\equiv - \sum_{n=1}^3V^{(n)}_1(t)T_i^{(n)}\frac{1}{a}(ip^i)HD_1(t)\delta_L(\bm{p})(2f(t) + 1) - \frac{1}{a}\sum_{n=1}^3\dot{V}^{(n)}_1T_i^{(n)}(ip^i)D_1(t)\delta_L(\bm{p}) \nonumber\\ 
&\quad + \sum_{n=1}^5V^{(n)}_2(t)\left\{H(f(t) + 1)T_{ij}^{(n)}\frac{1}{a}\frac{\partial}{\partial p^j}\left[p^iD_1(t)\delta_L(\bm{p})\right] + \frac{H}{a}\sum_{n=1}^5 V^{(n)}_2(t)T_{jk}^{(n)}  p^i\frac{\partial}{\partial p^k}\left[\frac{p^ip^j}{p^2}f(t)D_1(t)\delta_L(\bm{p})\right]\right. \nonumber\\
&\quad \left.- \frac{H}{a}\sum_{n=1}^5 V^{(n)}_2(t)T_{ij}^{(n)}\frac{p^ip^j}{p^2}f(t)D_1(t)\delta_L(\bm{p}) \right\} + \sum_{n=1}^5 \frac{1}{a}\dot{V}^{(n)}_2(t)T_{ij}^{(n)}\frac{\partial}{\partial p^j}\left[p^iD_1(t)\delta_L(\bm{p})\right]\nonumber\\
&\quad -  iHf(t)\sum_{n=1}^3\delta_1^{(n)}(t)T_i^{(n)}\left(\frac{\ddot{D}_1(t)}{\dot{D}_1(t)} + 2H \right)D_1(t)\left(\frac{p^i}{p^2}\delta_L(\bm{p}) - \frac{\partial}{\partial p^i}\left[\delta_L(\bm{p})\right]\right) \nonumber\\
&\quad -iH\sum_{n=1}^3\dot{\delta}_1^{(n)}(t)T_i^{(n)}\frac{p^i}{p^2}f(t)D_1(t)\delta_L(\bm{p}) + iH\sum_{n=1}^3\dot{\delta}_1^{(n)}(t)T_j^{(n)}\frac{\partial}{\partial p^j}\left[f(t)D_1(t)\delta_L(\bm{p})\right] \nonumber\\
&\quad +Hf(t)\left(\frac{\ddot{D}_1(t)}{\dot{D}_1(t)} + 2H \right)D_1(t)\sum_{n=1}^5 \delta_2^{(n)}(t)\left\{T_{ij}^{(n)}\frac{\partial}{\partial p^j}\left[\frac{p^i}{p^2}\delta_L(\bm{p})\right] - T_{jk}^{(n)}\frac{\partial}{\partial p^j}\frac{\partial}{\partial p^k}\left[\delta_L(\bm{p})\right] \right\} \nonumber\\
&\quad + \sum_{n=1}^5 H\dot{\delta}^{(n)}_2(t)T_{ij}^{(n)}\frac{\partial}{\partial p^j}\left[\frac{p^i}{p^2}f(t)D_1(t)\delta_L(\bm{p})\right] - \sum_{n=1}^5 H\dot{\delta}^{(n)}_2(t)T_{jk}^{(n)}\frac{\partial}{\partial p^j}\frac{\partial}{\partial p^k}\left[f(t)D_1(t)\delta_L(\bm{p})\right] \nonumber\\
&\quad  -\left\{3iH(f(t) +2)\frac{\partial}{\partial p^i}\left[D_1(t)\delta_L(\bm{p})\right] \right\}\sum_{n=1}^3\dot{\Phi}_1^{(n)}(t)T_i^{(n)}-iH\sum_{n=1}^3\dot{\Phi}_1^{(n)}(t)T_l^{(n)}p^i \frac{\partial}{\partial p^l_1}\left[\frac{p^i}{p^2}f(t)D_1(t)\delta_L(\bm{p})\right] \nonumber\\
&\quad - 3i\sum_{n=1}^3\ddot{\Phi}_1^{(n)}(t)T_i^{(n)}\frac{\partial}{\partial p^i}\left[D_1(t)\delta_L(\bm{p})\right]+ 3H(f(t) + 2)\sum_{n=1}^5 \dot{\Phi}^{(n)}_2(t)T_{ij}^{(n)} \frac{\partial}{\partial p^i}\frac{\partial}{\partial p^j}\left[D_1(t)\delta_L(\bm{p})\right]\nonumber\\
&\quad+iH\sum_{n=1}^5 \dot{\Phi}^{(n)}_2(t)T_{lm}^{(n)}p^i\frac{\partial}{\partial p^l}\frac{\partial}{\partial p^m}\left[\frac{p^i}{p^2}f(t)D_1(t)\delta_L(\bm{p})\right]+ 3\sum_{n=1}^5 \ddot{\Phi}^{(n)}_2(t)T_{ij}^{(n)} \frac{\partial}{\partial p^i}\frac{\partial}{\partial p^j}\left[D_1(t)\delta_L(\bm{p})\right],
\label{sourcetermfull}
\end{align}
where the subscripts ``$1$'' and ``$2$'' mean the dipole and quadrupole of the sh-mode, respectively, again. As we have explained in Sec.~\ref{subsec:sh-mode}, we neglected the quadrupole contribution of $\mathcal{O}(\kappa\epsilon^2)$ in our specific formulation for structure formation up to $\mathcal{O}(\kappa\epsilon)$; nevertheless, we list the $\mathcal{O}(\kappa\epsilon^2)$ terms for the convenience of possible extension of the formulation.

However, even if we use the generalized form in Eq.~(\ref{sourcetermfull}), for small scales $p/aH\gg1$, neglecting the terms that are the higher order infinitesimal of  
$\mathcal{O}(aH/p)$
starting from the third term in the second line of Eq.~(\ref{sourcetermfull}) will lead us to Eq.~(\ref{sourceterm}). 

\section{\label{AppendixB}Analytic Approximations}

In this appendix, we find the analytic approximate for the solutions of the equations and the sh-mode and then predicted power spectrum subsequently. 
First, we assume an initial period of matter dominance as $a\rightarrow 0 $ or $t\rightarrow 0$; then, the background matter density yields $\rho_0={\Omega_m H_0^2}/{a^3}$ from Eq.~(\ref{background1}).
With these, 
we will find the analytic approximation for the scalar field $\phi_0$.
Assuming the previous initial period in matter dominance as $t\rightarrow0$, we may rewrite Eq.~(\ref{background2}) 
using the relation $a(t)=(\frac{3}{2}H_0\sqrt{\Omega_m}t)^{2/3}$ as the ordinary differential equation in terms of $t$ or $a(t)$.
By defining $\tilde{t} \equiv H_0t$, 
Eq.~(\ref{background2}) leads to the solution of $\phi_0$ written as
\begin{align}
\phi_0 \equiv \bar{\phi}_0 \tilde{\phi}_0  = \bar{\phi}_0 F\frac{\sin{mt}}{mt} \simeq \bar{\phi}_0F(1 -\frac{\tilde{m}^2\tilde{t}^2}{6}),
\label{phi0ini}
\end{align}
where $F$ is the amplitude of the dimensionless background scalar field $\tilde{\phi_0}$, $\bar{\phi}_0$ a constant with the dimension of the scalar field $\phi$, and $\tilde{m} \equiv m/H_0$, and $\tilde{r} \equiv \frac{1}{6}({\overline{\phi}_0}/{M_{pl}})^2$ are the dimensionless parameters. From Eq.~(\ref{background3}), $F$ obeys
\begin{align}
1-\Omega_m = \tilde{r}\tilde{m}^2(\tilde{\phi}_0(F,a)|_{a=1})^2 + \tilde{r}\left({\partial{\tilde{\phi}_0(F,a)}\over \partial\tilde{t}}\bigg|_{a=1}\right)^2,
\label{eq:friedmann}
\end{align}
from which we can obtain the solution for $F$ making use of the initial condition $\phi_0(0)\rightarrow {\rm const}$.

Similar to the definition that $\phi_0 \equiv\overline{\phi}_0\tilde{\phi}_0$, we define parameters and nondimensionalized dipole perturbation as $\phi^{(n)}_1 \equiv \overline{\phi}_0\tilde{\phi}^{(n)}_1$, $\tilde{\delta}^{(n)}_1\equiv \frac{1}{H_0}\delta^{(n)}_1$,$\tilde{\Psi}^{(n)}_1\equiv \frac{1}{H_0}\Psi^{(n)}_1$,$\tilde{\Phi}^{(n)}_1\equiv \frac{1}{H_0}\Phi^{(n)}_1$. 
Then, the equations that the sh-mode follows read

\begin{gather}
\label{B18}
\tilde{\delta}^{(n)}_1 +3\tilde{\Phi}^{(n)}_1=0,\\
\label{B21}
\tilde{\Psi}^{(n)}_1 + \tilde{\Phi}^{(n)}_1 = 0,\\
\label{V'eq}
{\partial V'^{(n)}_1\over \partial\tilde{t}} - \tilde{\Psi}^{(n)}_1 =0,\\
\label{klein}
{\partial^2\tilde{\phi}^{(n)}_1\over\partial\tilde{t}^2}+ 3\frac{1}{a}{\partial a \over \partial\tilde{t}} {\partial\tilde{\phi}^{(n)}_1\over\partial\tilde{t}} + \tilde{m}^2 \tilde{\phi}^{(n)}_1 + \left(3{\partial\tilde{\Phi}^{(n)}_1\over\partial\tilde{t}} - {\partial\tilde{\Psi}^{(n)}_1\over\partial\tilde{t}}- 3\frac{1}{a}{\partial a \over \partial\tilde{t}} \tilde{\Psi}^{(n)}_1\right){\partial\tilde{\phi}_0\over\partial\tilde{t}} - 2\tilde{\Psi}^{(n)}_1{\partial^2\tilde{\phi}_0\over\partial\tilde{t}^2} =0,
\\
\label{B20}
2{\partial \tilde{\Phi}^{(n)}_1\over\partial\tilde{t}}-2\frac{1}{a}{\partial a \over \partial\tilde{t}}\tilde{\Psi}^{(n)}_1  =-3\Omega_ma^{-3}V'^{(n)}_1 -6\tilde{r}{\partial\tilde{\phi}_0\over\partial\tilde{t}}\tilde{\phi}^{(n)}_1.
\end{gather}
 
Next, we assume the following power-law time dependence for the perturbations in the limit of $t\to0$,
\begin{align}
\tilde{\delta}^{(n)}_1 &= e^{(n)}\mathcal{A}_{\rm I}\tilde{t}^{\alpha},\\
\tilde{\phi}^{(n)}_1 &= e^{(n)}(\mathcal{D} + \mathcal{D}_{\rm I}\tilde{t}^{\gamma}),
\end{align}
where 
$\mathcal{A}_{\rm I}$, $\mathcal{D}$ and $\mathcal{D}_{\rm I}$ are some constants to be determined subsequently. At the same time, we have $\tilde{\phi}_0 =F(1 -\frac{\tilde{m}^2\tilde{t}^2}{6})$ following Eq.~(\ref{phi0ini}).

Hence, combining Eq.~(\ref{B20}) with Eq.~(\ref{V'eq}), we find 
\begin{align}
 \left(-\frac{2}{3}\alpha^2 -\frac{10}{9}\alpha\right)\mathcal{A}_{\rm I}\tilde{t}^{\alpha-2}-6\tilde{r}F\tilde{m}^2\mathcal{D} -2\tilde{r}F\tilde{m}^2(\gamma + 2)\mathcal{D}_{\rm I}\tilde{t}^{\gamma}=0.
\end{align}
When $\tilde{t}\rightarrow 0$ with $\alpha,\gamma \neq0$, the only nontrivial solution is the case that $\alpha = 2$. Therefore, we obtain
\begin{align}
\label{A_1eq}
 22\mathcal{A}_{\rm I}+ 27\tilde{r}F\tilde{m}^2\mathcal{D} =0.
\end{align}
In a similar way,
Eq.~(\ref{klein}) reduces to
\begin{align}
  (\gamma^2+\gamma) \mathcal{D}_{\rm I}\tilde{t}^{\gamma-2} + \tilde{m}^2(\mathcal{D} + \mathcal{D}_{\rm I}\tilde{t}^{\gamma})  +\frac{4}{9}\mathcal{A}_{\rm I}(\alpha+1)F\tilde{m}^2\tilde{t}^{\alpha}  + \frac{2}{9}F\tilde{m}^2\mathcal{A}_{\rm I}\tilde{t}^{\alpha}=0.
\end{align}
Similarly, when $\tilde{t}\rightarrow 0$ with $\alpha,\gamma \neq0$, we have $\gamma=2$ so that
\begin{align}
\label{D_1eq}
6\mathcal{D}_{\rm I} + \tilde{m}^2\mathcal{D} =0.
\end{align}
Since we have defined $
e^{(n)}$ for the initial dipole direction of the sh-mode in Eq.~(\ref{defen}), solving for Eqs.~(\ref{A_1eq}) and (\ref{D_1eq}) by setting to $\mathcal{D} =1$ without loss of generality,
we obtain the analytic solutions as the approximates of $t\to 0(a\to 0)$,
\begin{align}
\phi^{(n)}_1 &\simeq H_0\bar{\phi}_0(1-\frac{1}{6}\tilde{m}^2\tilde{t}^2)e^{(n)},\\
\delta^{(n)}_1 &\simeq -\frac{27}{22}H_0\tilde{m}^2\tilde{r}F\tilde{t}^2e^{(n)},\\
V'^{(n)}_1 &\simeq -\frac{3}{22} \tilde{m}^2\tilde{r}F\tilde{t}^3e^{(n)}.
\label{Vsh'}
\end{align}
We write again the modified power spectrum $P(a,k,\theta)$ in Eqs.~(\ref{Pak}) and (\ref{defcalR}) so that
\begin{align}
P(a,\bm{k})=P(a,k,\theta)&=P_0(a,k)\left(1+ \epsilon^2 k^2\cos^2\theta {\cal R}(a)\right),\nonumber 
\\
\textrm{with} \quad {\cal R}(a) &=\frac{3}{25\pi}\frac{1}{\Omega_m H_0^2}
\left(\mathcal{I}(a)-\mathcal{J}(a)\right)^2,
\nonumber
\end{align}
following which we show a quantitative order estimation of the correction to LSS  power spectrum by applying the previous analytic approximates in the limit of  of matter dominance.
As we are interested in the evolution after early matter dominance era ($a_{eq} \ll a \ll 1$), the Hubble parameter can be approximated as (see Eq.~(42) of Ref.~\cite{nan2021dark})
\begin{align}
    \widetilde{H}(a)\equiv H(a)/H_0=\sqrt{\widetilde{r}\widetilde{m}^2\widetilde{\phi}_0^2+\Omega_m a^{-3} \over 1-\widetilde{r} a^2  \widetilde{\phi}_0'^2}\simeq \sqrt{\Omega_m a^{-3}}.
    \label{eq:ha2}
  \end{align}

Also, in the standard cosmological model, growth mode and decay mode become
\begin{align}
&D_1(a) = \frac{5\Omega_m}{2}\tilde{H}(a)\int^a_0\frac{\mathrm{d}a'}{(a'\tilde{H}(a'))^3}
\simeq a,\\
&D_2(a) = \frac{\tilde{H}(a)}{\sqrt{\Omega_m}}  
\simeq a^{-3/2}\quad .
\end{align}
By using this, we use the linear growth rate $f(a) \simeq 1$ during matter dominance.
Using these approximations, we estimate $\cal{I},\cal{J}$ as
\begin{align}
\mathcal{G}(a)&\simeq \frac{4}{99}{(1-\Omega_m)\over F (\Omega_m)^{3/2}}\left\{  \frac{13}{2}a^{\frac{9}{2}}\right\},\\
\mathcal{I}(a)&\simeq a^{-\frac{5}{2}}\int^a_0 da' a'
\mathcal{G}(a'),
\\ 
\mathcal{J}(a) &\simeq \int^a_0da' a'^{-\frac{3}{2}}\mathcal{G}(a').
\end{align}
On the other hand, from Eq.~(\ref{Vsh'}), recalling $V^{(n)}_1=-V'^{(n)}_1/a$ and using a general approximation for parameters $F^2\tilde{r}\tilde{m}^2\simeq1-\Omega_m$ that follows Eq.~(\ref{eq:friedmann}), we obtain
\begin{align}
V_1(a)&\simeq\frac{4}{99}F\tilde{r}\frac{\tilde{m}^2}{\Omega_m }\frac{ 1}{\sqrt{\Omega_m}}a^{\frac{7}{2}}\nonumber\\
&\simeq\frac{4}{99}{(1-\Omega_m)\over F (\Omega_m)^{3/2}}a^{\frac{7}{2}},
\end{align}
hence
\begin{align}
\mathcal{I}(a)&\simeq \frac{4}{99}{(1-\Omega_m)\over F (\Omega_m)^{3/2}}a^{4},
\\ 
\mathcal{J}(a) &\simeq \frac{13}{8}\mathcal{I}(a).
\end{align}

Now we can approximate $\mathcal{R}(a)$ in Eq.~(\ref{defcalR}) as
\begin{align}
{\mathcal R}(a) =\frac{3}{25\pi}\frac{1}{\Omega_m H_0^2}
\left(\frac{5}{8}\right)^2\left(\frac{4}{99}{(1-\Omega_m)\over F (\Omega_m)^{3/2}}\right)^2a^8.
\end{align}
Eventually, the approximate solution of the power spectrum becomes 
\begin{align}
P(a,k,\theta)\simeq P_0(a,k)\left(1+ \epsilon^2 k^2\cos^2\theta\frac{3}{4\pi}\frac{1}{\Omega_m H_0^2}
\left(\frac{1}{99}{(1-\Omega_m)\over F (\Omega_m)^{3/2}}\right)^2a^8\right).
\end{align}
We note that the accuracy of this analytic approximation for estimation is strongly restricted by the validity of approximations for $V_1$ in Eq.~(\ref{Vsh'}) and $f(a)\simeq1$ in the late time,
but it helps to understand the parameter dependence and time-evolution behavior of the model modification to the matter power spectrum. 

\bibliographystyle{apsrev4-2}
\bibliography{arxiv-resub}

\end{document}